\documentclass[journal=nalefd,manuscript=article,layout=twocolumn]{achemso}
\usepackage[version=3]{mhchem}
\usepackage{graphicx}
\usepackage{color}
\usepackage{amsmath}
\usepackage{amsfonts}
\usepackage{colortbl}
\usepackage{footnote}
\usepackage{amssymb}	
\usepackage{bm} 		
\usepackage{hyperref}
\usepackage{atbegshi}
\usepackage{float} 			
\usepackage[bottom]{footmisc}				
\usepackage{booktabs,caption}
\usepackage[flushleft]{threeparttable}
\usepackage{footnote}
\usepackage{multicol} 		
\usepackage{wrapfig}		
\usepackage{booktabs,caption}
\usepackage[flushleft]{threeparttable}

\author{Sebasti\'{a}n Castilla} 
\author{Bernat Terr\'{e}s} \affiliation{ICFO - Institut de Ci\`{e}ncies Fot\`{o}niques, The Barcelona Institute of Science and Technology, Castelldefels (Barcelona) 08860, Spain}
\author{Marta Autore} \affiliation{CIC nanoGUNE, E-20018 Donostia-San Sebastian, Spain}
\author{Leonardo Viti} \affiliation{NEST, CNR - Istituto Nanoscienze and Scuola Normale Superiore, 56127 Pisa, Italy}
\author{Jian Li} \affiliation{State Key Laboratory of Analytical Chemistry for Life Science, School of Chemistry and Chemical Engineering, Nanjing University, Nanjing 210023, China}
\author{Alexey Y. Nikitin} \affiliation{Donostia International Physics Center (DIPC), Donostia-San Sebastián, Spain}  \alsoaffiliation{IKERBASQUE, Basque Foundation for Science, 48013 Bilbao, Spain}
\author{Ioannis Vangelidis} \affiliation{Department of Materials Science and Engineering, University of Ioannina, GR-45110 Ioannina, Greece}
\author{Kenji Watanabe} \affiliation{Advanced Materials Laboratory, National Institute for Material Science, 305-0044, Tsukuba, Japan}
\author{Takashi Taniguchi} \affiliation{Advanced Materials Laboratory, National Institute for Material Science, 305-0044, Tsukuba, Japan}
\author{Elefterios Lidorikis} \affiliation{Department of Materials Science and Engineering, University of Ioannina, GR-45110 Ioannina, Greece}
\author{Miriam S. Vitiello} \affiliation{NEST, CNR - Istituto Nanoscienze and Scuola Normale Superiore, 56127 Pisa, Italy}
\author{Rainer Hillenbrand} \affiliation{CIC nanoGUNE, E-20018 Donostia-San Sebastian, Spain} \alsoaffiliation{IKERBASQUE, Basque Foundation for Science, 48013 Bilbao, Spain}
\author{Klaas-Jan Tielrooij} \email{klaas.tielrooij@icn2.cat}  \affiliation{ICFO - Institut de Ci\`{e}ncies Fot\`{o}niques, The Barcelona Institute of Science and Technology, Castelldefels (Barcelona) 08860, Spain} \alsoaffiliation{Current address: Catalan Institute of Nanoscience and Nanotechnology (ICN2), Barcelona Institute of Science and Technology, Campus UAB, Bellaterra, Barcelona, 08193, Spain} 
\author{Frank H.L. Koppens} \email{frank.koppens@icfo.eu} \affiliation{ICFO - Institut de Ci\`{e}ncies Fot\`{o}niques, The Barcelona Institute of Science and Technology, Castelldefels (Barcelona) 08860, Spain} \alsoaffiliation{ICREA - Instituci\'o Catalana de Re\c{c}erca i Estudis Avancats, 08010 Barcelona, Spain}

\title{Fast and sensitive terahertz detection using an antenna-integrated graphene $pn$-junction}

\begin{document}
\setcounter{tocdepth}{1} 
\setcounter{page}{1}

\twocolumn[
  \begin{@twocolumnfalse}
    \maketitle

\begin{abstract}
Although the detection of light at terahertz (THz) frequencies is important for a large range of applications, current detectors typically have several disadvantages in terms of sensitivity, speed, operating temperature, and spectral range. Here, we use graphene as photoactive material to overcome all of these limitations in one device. We introduce a novel detector for terahertz radiation that exploits the photo-thermoelectric effect, based on a design that employs a dual-gated, dipolar antenna with a gap of $\sim$100 nm. This narrow-gap antenna simultaneously creates a $pn$-junction in a graphene channel located above the antenna, and strongly concentrates the incoming radiation at this $pn$-junction, where the photoresponse is created. We demonstrate that this novel detector has excellent sensitivity, with a noise-equivalent power of 80 pW/$\sqrt{\rm Hz}$ at room temperature, a response time below 30 ns (setup-limited), a high dynamic range (linear power dependence over more than 3 orders of magnitude) and broadband operation (measured range 1.8 -- 4.2 THz, antenna-limited), which fulfills a combination that is currently missing in the state of the art. Importantly, based on the agreement we obtain between experiment, analytical model, and numerical simulations, we have reached a solid understanding of how the PTE effect gives rise to a THz-induced photoresponse, which is very valuable for further detector optimization. 
\end{abstract}
  \end{@twocolumnfalse}
  ]

\maketitle

Photodetectors operating at THz frequencies play an important role in many applications in the fields of medicine, security, quality testing, chemical spectroscopy and more \cite{tonouchi07, ferguson2002a, lee07, mittleman03, appleby07, federici05, siegel04}. One of the main benefits of THz radiation is its non-invasive nature and its capability to penetrate most dielectric materials, which are typically opaque at non-THz frequencies. For example, in the case of medical imaging and security applications, THz radiation offers clear advantages since it is non-ionizing due to its low photon energy (in the meV range) in contrast with conventional X-ray radiation with much higher photon energy (in the keV range), leading to strongly reduced health risks. Furthermore, THz detectors are expected to play an enabling role for data communication at THz bit rates \cite{Kleine-Ostmann2011, Ma2017, mittendorff17}. For many of these applications, the ideal THz detector would meet the following five requirements: it should be highly sensitive ($i.e.$\ have a low noise-equivalent power, NEP), operate at room temperature, give a fast photoresponse, have a high dynamic range (the range between the lowest and highest measurable incident light power), and work over a broad range of THz frequencies, in particular above 1 THz. 
\\

Commercially available room-temperature THz detectors, for example pyroelectric detectors and Golay cells, are reasonably sensitive with an NEP on the order of $\sim$1 nW/$\sqrt{\rm Hz}$. However, their response time is very long: 100 and 30 ms, respectively \cite{gentec, tydex}. Bolometric THz detectors, on the other hand, can be highly sensitive with an NEP  of $\sim$0.5 pW/$\sqrt{\rm Hz}$, while simultaneously showing fast operation with a response time of $\sim$50 ps. However, these detectors require cryogenic temperatures ($\sim$4 K) and suffer from a narrow dynamic range (maximum detectable power $\sim$0.1 $\mu$W) \cite{scontel}. Schottky diodes, although combining high speed (response time in the picoseconds regime) and high sensitivity (NEP of 10 - 100 pW/$\sqrt{\rm Hz}$), have a low frequency cut-off (operation only below $\sim$1 THz) and a small dynamic range \cite{sizov10, vdiodes}. Thus, currently there are no commercially available THz detectors that simultaneously meet all five requirements. 
\\

Owing to its exceptional optoelectronic properties and broadband absorption spectrum (from the visible down to the GHz-THz range) graphene is a highly interesting photoactive material for detecting light \cite{low14, basov08, koppens2014, Bonaccorso2010b, tredicucci12, mittendorff13}. During the past couple of years, there were several experimental demonstrations of graphene-based photodetection in the GHz-THz range. These detectors were based on various operating mechanisms. First of all, there were reports describing plasma wave-assisted THz detection, typically in the overdamped regime \cite{tredicucci12, tredicucci14, chalmers14, chalmers17, bandurin18}. This scheme has resulted in high sensitivities, with an NEP below nW/$\sqrt{\rm Hz}$, but typically this mechanism is reported for radiation below 1 THz. Secondly, ballistic graphene rectifiers were demonstrated with excellent sensitivity, but only operating below 1 THz \cite{manchester17}. Moreover, a graphene-antenna coupled bolometer for detecting GHz radiation was shown with promising values of sensitivity at low temperatures \cite{chalmers18}.
\\

Alternatively, one can exploit the photo-thermoelectric (PTE) effect, where absorbed THz light heats up the graphene electrons\cite{mics15}, subsequently creating an electron-heat driven photoresponse if an asymmetry is present in the device \cite{song11}. Such an asymmetry could be created, for example, by using two different contact metals or by using two adjacent graphene regions of different doping, $e.g.$\ forming a junction. Photodetection based on the PTE effect in graphene was first shown for visible light, where \textit{interband} absorption of light occurs \cite{lemme2011, gabor2011d}. More recently, also photoresponses in the THz frequency regime, where absorption occurs through \textit{intraband} processes, were attributed to the PTE effect \cite{fuhrer14, pte18}. Moreover, several of the previously mentioned studies exploiting alternative mechanisms also attributed a potentially significant fraction of the observed THz photoresponse to the photo-thermoelectric effect \cite{chalmers14, chalmers17, tredicucci12, tredicucci14, manchester17, bandurin18, mittendorff13}. Clearly, some controversy exist on which photoresponse mechanism dominates for graphene excited by THz light, which has hampered the development of more optimized detectors. Furthermore, the main challenge for exploiting the PTE effect for THz detection is the large mismatch between the large area of the incoming radiation and the small photo-active area of graphene, where the PTE effect occurs.
\\


\begin{figure*} [h!!!!!]
   \centering
   \includegraphics [scale=0.4]
   {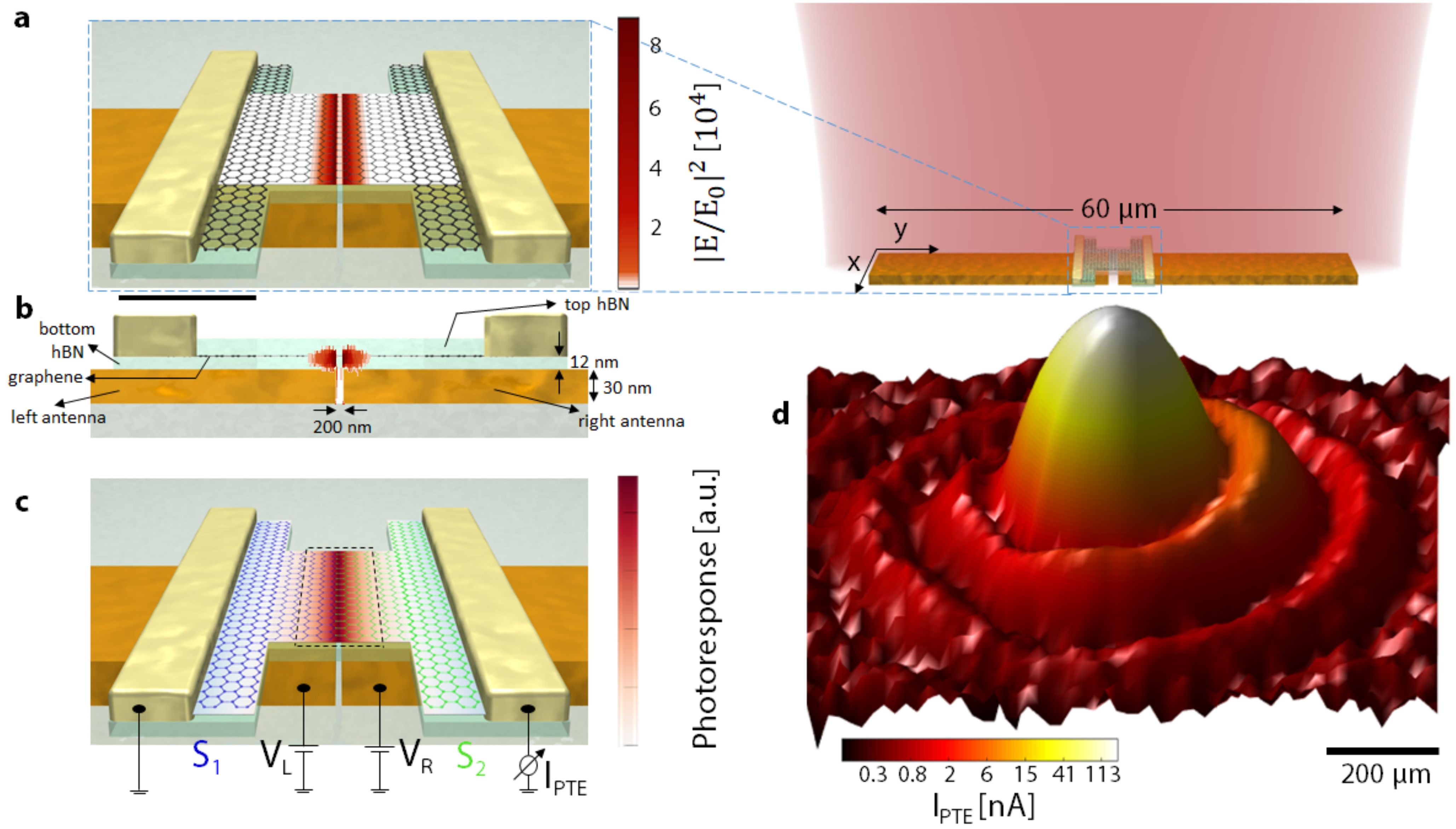}
   \caption{ 
\footnotesize \textbf{a)} Schematic representation (right; not to scale) of the antenna-integrated $pn$-junction device and a zoom of the central part of the THz PTE detector (left; to scale), consisting of an ``H-shaped'' graphene channel, contacted by source and drain electrodes. Underneath the graphene channel, there are two antenna branches that concentrate the incident THz light around the antenna gap region. The color map superimposed on the device shows the simulated power profile ($|E/E_0|^2$, where $E_0$ is the incident electric field) at a position 5 nm below the graphene channel. The black scale bar corresponds to 1.6 $\mu$m. 
\textbf{b)} Side view of the device design, with the superimposed color map again indicating the normalized power profile as in panel \textbf{a}. The region where the field is strongly enhanced by the antenna overlaps with the central part of the graphene channel. 
\textbf{c)} Same as panel \textbf{a}, now indicating how the antenna branches serve as local gates by applying voltages $V_{\rm L}$ and $V_{\rm R}$. Appropriate voltages will create a $pn$-junction in the central part of the graphene channel, directly above the antenna gap (which is where incident THz light is concentrated by the antenna). The color map superimposed on the device is a simulation that shows the photoresponse created by local photoexcitation, varying the position of photoexcitation (see also SI). The largest photoresponse is created when photoexcitation occurs around the junction region. The photoresponse then decreases exponentially when moving away on both sides from the junction, with the exponential decay length given by the cooling length $\ell_{\rm cool}$. The photoactive area (dashed rectangle) therefore has a length $2\cdot\ell_{\rm cool}$, and a width $w$, which is the width of the central part of the graphene channel. 
\textbf{d)} Photocurrent image (log-scale) obtained by scanning the detector in the focal plane of a focused laser beam at 3.4 THz. We use our QCL with an average power of 84.1 $\mu$W, and a peak irradiance in the center of the focus of 1200 W/m$^2$. The THz light is polarized parallel to the antenna axis. The observation of the Airy pattern with multiple observable rings indicates excellent detector sensitivity. 
}
   \end{figure*}


Here, we solve this issue by introducing an antenna-integrated THz photodetector, based on high-mobility, gate-tunable, hexagonal BN (hBN)-encapsulated graphene, where the incoming THz radiation is concentrated such that it overlaps with the small photoactive area of the graphene. Using the gate-tunability of the detector, we find that the PTE effect is the dominant photoresponse mechanism. We support this with a quantitative comparison of the device response with numerical simulations and an analytical model of the PTE photoresponse. We furthermore show that, owing to its novel device design, our PTE THz photodetector meets all five requirements of an ideal detector. In addition, it has the advantage of being based on low-cost materials with scalable integration capabilities with the well established CMOS electronics for low-cost imaging systems \cite{Goossens2017}. Finally, it is very low in power consumption, as it is a passive device.
\\

In the following, we will first explain how our antenna-integrated $pn$-junction THz detector works, followed by the experimental characterization of the detector. Subsequently, we provide an analytical model of the PTE detector and numerical simulations of the absorption enhancement of graphene induced by the antenna, and compare these results with the experiments. Finally, we will compare the THz photodection performance to the state of the art. 
\\

Our THz photodetector is based on a novel design (see Fig.\ 1a-c), which works as follows. The detector contains a dipole antenna that is located $\sim$15 nm below a graphene channel. The antenna consists of two branches that are separated by a very narrow gap, with a size on the order of 100 nm. This antenna gap serves for focusing the incoming THz radiation into a very small spot in the graphene channel. Here, the concentrated field of the antenna leads to (intraband) absorption and the subsequent creation of hot carriers \cite{mics15}. Since the creation of a photoresponse from hot carriers requires a gradient in the Seebeck coefficient, we use the antenna branches simultaneously as split gates. We apply appropriate voltages ($V_{\rm L}$ and $V_{\rm R}$) to the left and right antenna branch, and through capacitive coupling this creates a $pn$-junction in the graphene channel, and thereby a THz-induced photoresponse. Thus, the antenna simultaneously creates the photoactive area in the graphene channel (located around the $pn$-junction, see Fig.\ 1c) and funnels incident radiation to this photoactive area, due to the very strong field enhancement of incident THz radiation above the gap between the two antenna branches (see Fig.\ 1a-b). 
\\

Compared to previous antenna-integrated, graphene-based THz detectors \cite{chalmers14, chalmers17, tredicucci12, tredicucci14, manchester17, bandurin18, mittendorff13, chalmers18}, the advantage of our design is that the antenna gap is much smaller (100 nm $vs.$ several microns), which means that the THz intensity is greatly enhanced ($\sim$4 orders of magnitude, see Fig.\ 1a). Also, there is no direct electrical connection between the antenna and the graphene, which means that there is no need for impedance matching to assure current flow between antenna and graphene. The fact that we simultaneously use our antenna for focusing light and as split gate, has the advantage that there automatically is very good overlap between the region where the incoming THz radiation is focused and the photoactive region of the graphene channel (see Fig.\ 1a-c). Furthermore, we use hBN-encapsulated graphene, which leads to graphene with high mobility and low residual doping. This means that the resistance of the graphene channel will be low, and we can tune the system relatively close to the Dirac point (point of lowest carrier density), where the Seebeck coefficient is largest. Finally,  we pattern the graphene channel into an ``H-shape'' with a relatively narrow (micron-sized) width. The narrow width of the central part of the channel leads to an enhanced photoresponse, as the hot carriers will have a higher temperature. The wider sides of the graphene channel reduce overall device resistance by minimizing contact resistance.
\\

We have fabricated two ``H-shaped'', high-mobility, hBN-encapsulated graphene devices with a dipolar antenna/gating structure, (see Methods and Supporting Information for details and optical pictures of the devices). We mainly show the results obtained from THz photodetector Device A and will mention some results from THz photodetector Device B. Both Device A and Device B have a width of the central part of the graphene channel (at the junction) of $w$ = 2 $\mu$m, whereas the gap sizes of the dipolar antennas are 200 and 100 nm, respectively. The vertical distance between antenna and graphene is given by the thickness of the bottom hBN layer, typically $\sim$15 nm, and thus small enough to warrant sufficient overlap between the electric field profile around the antenna gap, and the graphene (see also Fig.\ 1b). The graphene mobility for both devices is on the order of 20,000 cm$^2$/Vs, which is a lower bound as it is determined from two-terminal measurements (see Supporting Information, SI). We characterize the performance of the THz photodetector devices using two different setups, both containing a THz laser and optical components to focus the light at our THz detector. One setup contains a pulsed quantum cascade laser (QCL) operating at 3.4 THz, and two Picarin (tsupurica) lenses to focus the light; the other setup contains a continuous wave THz gas laser with tunable output frequency, and a parabolic mirror to focus the light (see Methods for details). The THz light is usually modulated by an optical chopper and the generated photocurrent is measured using a pre-amplifier and/or lock-in amplifier. We use typical incident THz powers in the range of several microwatts to several milliwatts.
\\


\begin{figure*} [h!!!!!]
   \centering
   \includegraphics [scale=0.5]
   {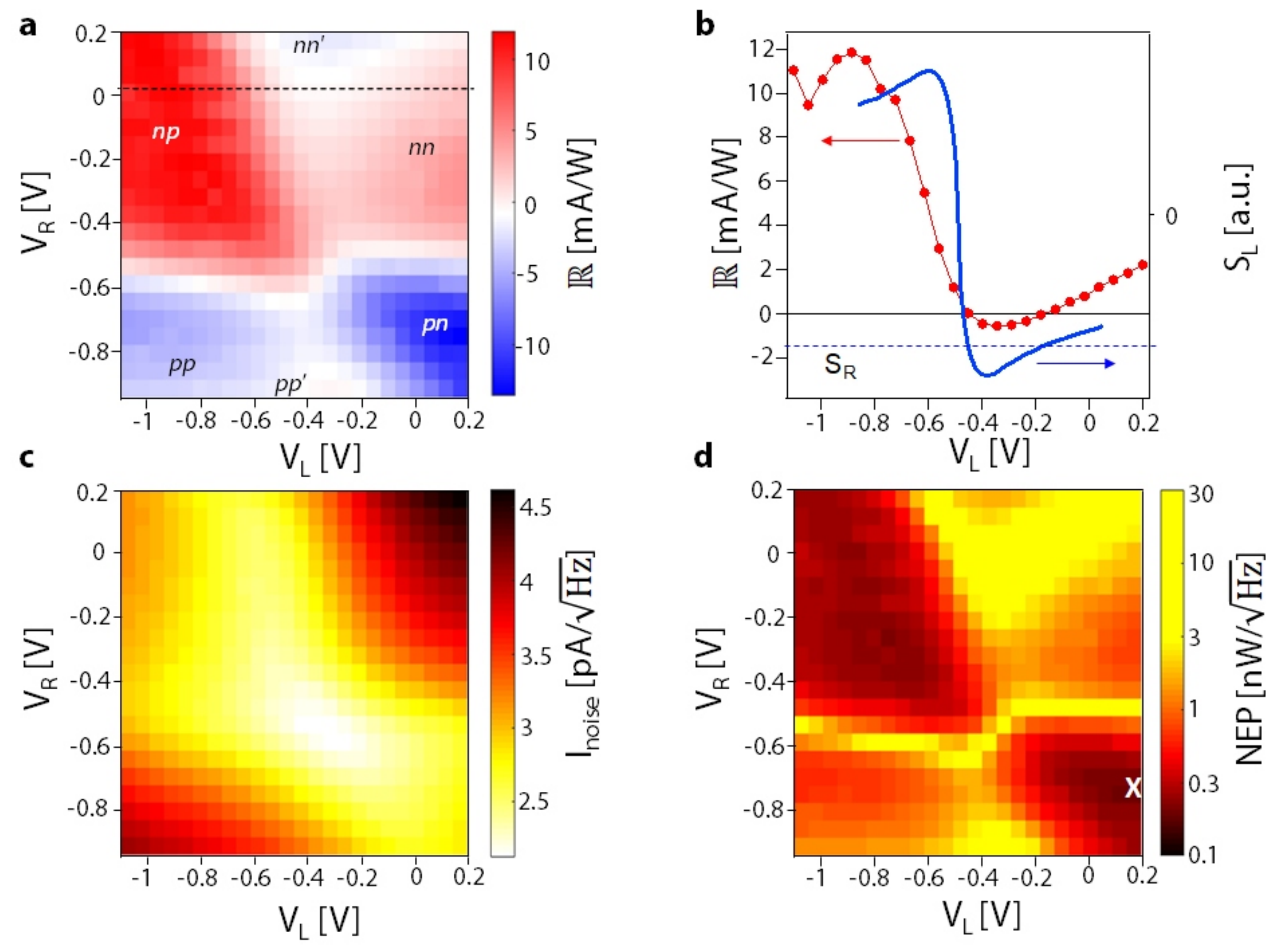}
   \caption{
\footnotesize \textbf{a)} Photoresponse as a function of voltages applied to the two antenna branches/gates, with radiation at 2.52 THz. We use our THz gas laser with an average incident power of $P_{\rm in}$ = 5 mW, and a peak irradiance in the center of the focus of 1.8$\cdot$10$^4$ W/m$^2$. The THz light is polarized parallel to the antenna axis. The sixfold pattern, as in Refs.\ \cite{gabor2011d, song11}, indicates that the photoresponse is generated through the PTE effect. The photoresponse is the photocurrent normalized by the power in a diffraction-limited spot, $i.e.$\ the responsivity $\mathbb{R}$. The maximum responsivity occurs in the $pn$- and $np$-regions. 
\textbf{b)} Line cut at the location of the dashed line in panel a, showing a double sign change (red dots and line; left vertical axis) as a function of carrier density (controlled through gate voltage $V_{\rm L}$). The blue line represents the Seebeck coefficient (calculated from the experimentally obtained graphene mobility, see SI; right vertical axis). The double sign change occurs due to the non-monotonous dependence of the Seebeck coefficient on carrier density: for a constant Seebeck coefficient in one region (dashed horizontal line), the Seebeck coefficient of the other region is first higher, then lower and then again higher, giving rise to two sign changes, as indeed observed experimentally. 
\textbf{c)} The extracted Johnson noise current, calculated from the resistance that was measured simultaneously with the result in panel \textbf{a}. 
\textbf{d)} The noise-equivalent power (NEP), extracted from the results in panels \textbf{a} and \textbf{c}, normalized to a diffraction-limited spot. The white cross indicates the gate configuration that corresponds to the lowest NEP: the left (right) gate at 0.20 V (-0.72 V), corresponding to an electron density of 7.5$\cdot10^{11}$ cm$^{-2}$, $E_{\rm F}$ = $+$100 meV (hole density of 3.6$\cdot10^{11}$ cm$^{-2}$, $E_{\rm F}$ = $-$70 meV). 
 }
   \end{figure*}


We first scan our photodetector (Device A) across the THz focus using motorized stages (in the QCL setup, see Methods). The dipolar antenna, with a length of 60 $\mu$m, is smaller than the THz focus (FWHM $\sim$200 $\mu$m), allowing us to spatially map out the intensity of the THz focus through the photocurrent $I_{\rm PTE}$. The results show a clear Airy pattern (see Fig.\ 1d), where we are able to observe several rings of the diffracted beam pattern. This suggests that our THz photodetector is very sensitive, considering that these rings contain only a very small fraction of total incident power of the THz beam ($P_{\rm in}$ = 84.1 $\mu$W). 
\\

Before proceeding with quantifying the sensitivity, we first exploit the gate-tunability to identify the photocurrent generation mechanism and determine the optimal operating point of our THz detector, by mapping out the photoresponse as a function of gate voltages $V_{\rm L}$ and $V_{\rm R}$ (see Fig.\ 2a). These measurements were done with the THz gas laser at 2.52 THz, and using Device A. The gate voltages independently control the carrier density (Fermi energy) of the two graphene regions and therefore the Seebeck coefficients $S_1$ and $S_2$. The Seebeck coefficient of graphene has a non-monotonous dependence on carrier density, where it first increases upon approaching the Dirac point and then changes sign when crossing the Dirac point, $i.e.$\ when going from hole to electron doping or vice versa (see Fig.\ 2b). Since the generated photocurrent $I_{\rm PTE} \propto (S_1 - S_2)$, this leads to the characteristic sixfold pattern, first shown in Ref.\ \cite{gabor2011d} for visible light, and explained in Fig.\ 2b. The fact that we also observe a sixfold pattern strongly suggest that our THz photoresponse is dominated by the PTE effect. To further confirm that the PTE mechanism dominates over alternative photocurrent mechanisms, such as bolometric and photogating effects, we measured the photocurrent as a function of bias voltage applied between the source and drain contacts. The drain current increases linearly with applied bias voltage, whereas the photocurrent remains constant (see SI), in contradiction with what is expected for the bolometric and photogating effects. Thus, these results show that the photo-thermoelectric effect is responsible for the observed THz photoresponse. We find the largest photoresponse in the $pn$-junction and $np$-junction regimes, as expected, relatively close to the Dirac point.
\\

We now proceed with quantifying the sensitivity of our THz photodetector. First, we identify the largest responsivity $\mathbb{R} = I_{\rm PTE}/P_{\rm diff}$ at the optimal gate configuration for Device A. We note that the responsivity that we use is the responsivity \textit{normalized by the power in a diffraction-limited spot with NA = 1}. In some of the literature, a responsivity is provided that is normalized by the amount of power that is actually incident in the experiment $P_{\rm in}$. However in the case where the wavelength is larger than the photodetector device this number does not only depend on the device performance, but also on how well the THz light is focused. Alternatively, one can use the responsivity normalized to the power that is absorbed in the actual graphene channel or in the photoactive area, which would yield an artificially high number \cite{fuhrer14}, as it is impossible to focus the THz light in such a small area. Here we choose the responsivity normalized by the incident power in a diffraction-limited spot, because this is arguably the most technologically relevant number (as this represents what will be reached when combining the detector with an optimized focusing system, such as using a silicon hemispherical lens \cite{chalmers14, chalmers17, bandurin18}), and it is the convention that is most commonly used in the literature on THz photodetection (see also Table 1). We calculate the power in a diffraction-limited spot using $P_{\rm diff} = P_{\rm in}\cdot A_{\rm diff}/A_{\rm focus}$, where $P_{\rm in}$ is the measured total incident THz power, $A_{\rm diff}$ is the calculated area of a diffraction-limited spot and $A_{\rm focus}$ is the measured area of the focused THz beam. Typically, we have $A_{\rm diff}/A_{\rm focus} \approx$ 1/60 (corresponding to a NA of $\sim$0.13 for our focusing system based on a parabolic mirror, see Methods for details). From Fig.\ 2a we extract a maximum responsivity of $\mathbb{R}$ = 14 mA/W (32 V/W). For Device B, we find a maximum responsivity value of $\mathbb{R}$ = 25 mA/W (105 V/W) (see Methods and SI). In both cases the THz light was at 2.52 THz (corresponding to a wavelength of 118.96 $\mu$m, 84 cm$^{-1}$). 
\\


\begin{figure} [H]
   \includegraphics [scale=0.6]
   {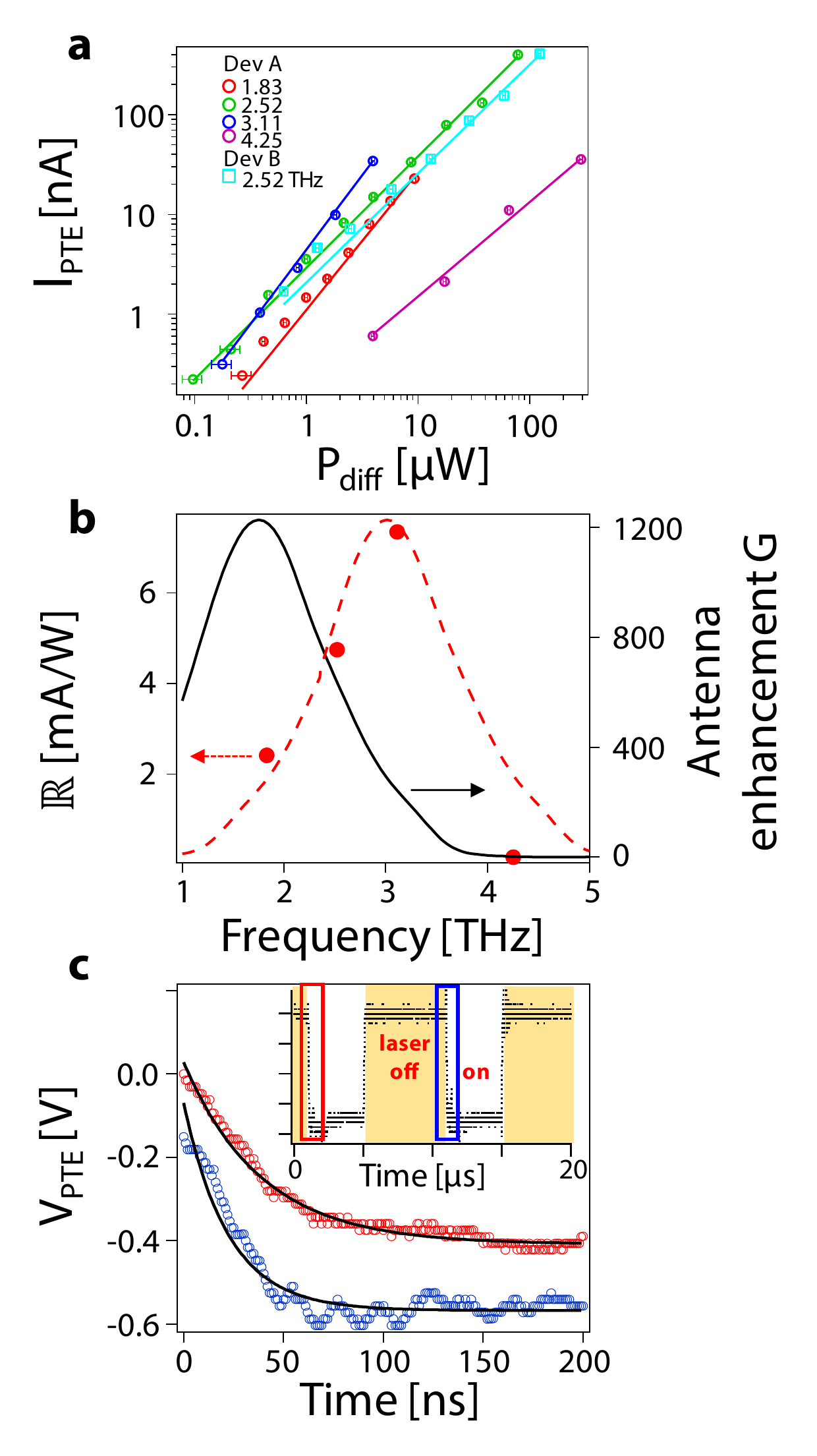}
   \caption{
\footnotesize \textbf{a)} Photocurrent as a function of THz laser power (in a diffraction-limited spot) $P_{\rm diff}$ in log-log scale. All round data points correspond to Device A, whereas the light blue squares correspond to Device B. The lines through the experimental data points are fits according to $I_{\rm PTE} \propto P_{\rm diff}^\gamma$. The obtained exponent is close to 1 for all data sets. For Device A, these data correspond to a gate configuration of $V_{\rm L} =$ 0 V and $V_{\rm R} =$ -0.67 V, corresponding to an electron (hole) density of 4.2$\cdot10^{11}$ (-2.7$\cdot10^{11}$) cm$^{-2}$. 
\textbf{b)} Responsivity $\mathbb{R}$ as a function of THz wavelength (red dots, left vertical axis), with the same (sub-optimal) gate configuration as in panel \textbf{a}. The black line (right vertical axis) shows the result for the antenna-induced absorption enhancement in the graphene channel. The red dashed line illustrates the trend of the experimental points.
\textbf{c)} Results of the pulsed laser experiment, where the photocurrent was amplified by a fast current pre-amplifer (Femto) and the data were acquired with a fast oscilloscope. The inset shows how the photovoltage $V_{\rm PTE}$ follows the switching of the pulsed laser. The red and blue (plotted with an offset) open dots show the obtained photovoltage in a small time window marked in the inset, with the black line giving the result of exponential fits with timescales of 40 (24) ns for the red (blue) curve. 
 }
   \end{figure}

Using the extracted responsivity, we now determine the sensitivity of the detector. For this, we note that our detector operates without bias, which means that it is limited by the Johnson or thermal noise, given by: $I_{\rm noise} = \sqrt{\frac{4k_{\rm B}T\Delta f}{R}}$. Here, $k_{\rm B}$ is the Boltzmann constant, $T$ is the temperature of operation, $\Delta f$ is the spectral bandwidth and $R$ the resistance. We show the Johnson noise current of Device A as a function of the two gates voltages in Fig.\ 2c, calculated from the measured resistance $R$. The resistance was measured simultaneously with the photoresponse in Fig.\ 2a, which we used to determine the responsivity $\mathbb{R}$. As expected, we see that a higher noise current occurs for lower resistance values (away from the Dirac point), whereas we obtain lower noise values closer to the Dirac point, where the graphene resistance is higher. The photodetection sensitivity is given by NEP$ = I_{\rm noise}/\mathbb{R}$, which we show as a function of the two gate voltages in Fig.\ 2d. The lowest values for the NEP occur at the $pn$- and $np$-regions close to the Dirac point, where the responsivity is highest and the noise is lowest. The lowest value of the NEP map was 200 pW/$\sqrt{\rm Hz}$ for Device A and 80 pW/$\sqrt{\rm Hz}$ for Device B (see Methods and SI).
\\

An important characteristic of an ideal THz detector is having a large range of powers over which the response is linear, $i.e.$\ a large dynamic range. Thus, we measure the photocurrent $vs.$ $P_{\rm diff}$ for four different THz frequencies as shown in Fig.\ 3a. We vary the THz power over more than 3 orders of magnitude (using the THz gas laser setup), and fit the data with a simple power law $I_{\rm PTE} \propto P_{\rm diff}^\gamma$. We obtain for Device A at 2.5 THz a power of $\gamma = 1.1 \pm 0.2$ and for Device B $\gamma = 1.1 \pm 0.15$ (95\% confidence intervals). This shows that the photoresponse depends linearly on the THz power over a range of more than three orders of magnitude. The reason for the linear photoresponse as a function of power is the fact that the photodetector operates in the weak heating regime, where $\Delta T \ll T_{\rm ambient}$, $i.e.$\ the change in temperature of the electronic system is smaller than the ambient temperature $T_{\rm ambient}$. When $\Delta T$ approaches $T_{\rm ambient}$, we expect a sub-linear dependence of photocurrent on power, with an exponent that tends to $\gamma$ = 0.5. 
\\

Regarding the range of frequencies where our detectors operate, we note that this is only limited by the antenna structure. The reason behind the spectrally ultra-broad photoresponse of graphene is the efficient heating of the electrons, which occurs irrespective of the wavelength of the incident light, $i.e.$\ whether intraband or interband light absorption occurs \cite{mics15, Tielrooij2015g}. We characterize the spectral response of our detector by measuring the responsivity, while varying the frequency from 1.83 to 4.25 THz (see Fig.\ 3b). We observe a trend where the responsivity peaks around 3 THz. This corresponds reasonably well with the antenna being optimized for a frequency of 2 THz using full wave simulations (see Methods and Fig.\ 3b). The discrepancy likely comes from the fact that a simplified structure was simulated, which didn't contain all the metallic parts that the actual device has. Importantly, these results confirm that the spectral range where the THz detector operates is currently limited only by the antenna. Thus, using more spectrally broad antennas or a combination of antennas one could extend the spectral range of our photodetector, covering the spectrum all the way from the ultraviolet, through the visible and infrared to the terahertz. \\

In the following, we discuss the speed of our PTE THz photodetector. We analyze the speed in Fig.\ 3c, where we rapidly switch the THz radiation on (white) and off (yellow) using our pulsed THz QCL (see Methods). We observe that the photoresponse of our detector $V_{\rm PTE}$ closely follows the laser switching behavior. We quantify the detector speed by fitting the $V_{\rm PTE}$ rise and fall dynamics (see also SI) with simple exponential equations, and obtain an exponential (1/e) response time of 32$\pm$11 ns, corresponding to a bandwidth of $\sim$5 MHz. In this measurement, the speed is most likely limited by the measurement electronics, namely by the 3.5 MHz bandwidth of the current pre-amplifier, rather than by the PTE THz detector itself. Indeed, the intrinsic speed of the detector is expected to be significantly higher, since ultimately the intrinsic response time is limited by the RC-time of the detector. Performing this calculation (see Methods), we obtain a rise time of 56 ps, corresponding to a detection rate of 6 GHz for our device with a mobility of 20,000 cm$^2$/Vs. For a device with a mobility of 100,000 cm$^2$/Vs (see SI), we find 9 ps, corresponding to 40 GHz. Thus, extremely fast THz photodetection with switching times in the picosecond range should be possible.
\\

We now discuss more of the underlying physics of the PTE detector, using a simple analytical model that provides the rationale behind our detector design. Owing to the difference in Seebeck coefficients at the $pn$-junction, a local photo-thermoelectric voltage is created, which leads to the flow of a photocurrent between the source and drain contacts that are connected to the graphene channel. The PTE photocurrent is then given by \cite{song11}
\begin{equation}
I_{PTE} = \frac{(S_{1}-S_{2})\,\Delta T}{R} \hspace{2mm}, 
\label{PC}
\end{equation}
where $S_1$ and $S_2$ are the Seebeck coefficients (also called thermopower) of the two regions of the graphene channel that are independently controlled by the gates/antenna branches, $\Delta T$ is the temperature increase of the electronic system induced by THz radiation, and $R$ is the total electrical resistance, accounting for the graphene and contact resistances. It is worth mentioning that this current is generated under zero applied source-drain bias voltage, resulting in very low detector noise (Johnson noise) and extremely low power consumption.
\\

Graphene is an ideal material to exploit the PTE effect for THz detection, because the term $(S_{1}-S_{2})\,\Delta T$ can be large and $R$ is typically small, in particular for high-quality, hBN-encapsulated graphene. The Seebeck coefficient of graphene is intrinsically quite large, on the order of 100 $\mu$V/K \cite{kim09} and $S_1$ and $S_2$ are independently tunable through the gates, meaning that $(S_{1}-S_{2})$ can be maximized. Furthermore, $\Delta T$ can be large in graphene (up to several thousand K), because of efficient heating of the electrons after absorbing THz light, due to strong electron-electron interactions, and because the hot carriers are relatively weakly coupled to the crystal lattice \cite{mics15}. 
\\

Our photodetector design maximizes the PTE THz photoresponse, particularly by maximizing $\Delta T$ and minimizing $R$. From a simple heat equation, the temperature increase $\Delta T$ (averaged over all charge carriers in the photoactive area) is given by 

\begin{equation}
\Delta T \approx \frac{P_{\rm abs}}{A_{\rm active}\Gamma_{\rm cool}} 
\label{deltaT}
\end{equation}

where $P_{\rm abs}$ is the amount of THz power that is absorbed in the active area of the graphene channel and $\Gamma_{\rm cool}$ is the heat conductivity that describes the coupling of the heated electron systems to its environment. The photoactive area is given by $A_{\rm active} = 2\ell_{\rm cool}\cdot w$, where $\ell_{\rm cool}$ is the hot-carrier cooling length, which can be seen as the length scale over which hot carriers can move before cooling down (typically 0.5 -- 1 $\mu$m at room temperature \cite{song11, gabor2011d, Tielrooij2015g, Tielrooij2018, macdonald09}, see also Fig.\ 1c). In the case of hBN-encapsulated graphene, $\Gamma_{\rm cool}$ is the out-of-plane, interfacial heat conductivity where hot graphene carriers couple to hyperbolic hBN phonons \cite{Tielrooij2018}. We optimize $\Delta T$ by maximizing $P_{\rm abs}$ and minimizing $A_{\rm active}$. We maximize the amount of absorbed THz power $P_{\rm abs}$ by using a dipole antenna with a narrow gap, which focuses the incoming THz radiation down to the small (compared to the THz radiation wavelength) graphene photoactive region. We further maximize $\Delta T$ by using a narrow channel width $w$ of 2 $\mu$m. Basically, the smaller the area where the incident power is absorbed, the smaller the amount of electrons that will share the heat, and therefore the larger the increase in temperature of the electronic system, $\Delta T$. 
\\

In order to further increase the responsivity, we reduce the overall resistance $R$ of the device, and optimize the shape of the graphene channel. We achieve low $R$ by using high-quality hBN-encapsulated graphene (see Fig.\ 1b) \cite{Wang2013g,Pizzocchero2016}. This method enables mobility values as high as 100,000 cm$^2$/Vs at room temperature, and low levels of intrinsic doping (see SI). We furthermore pattern the graphene channel in an ``H-shape'' (see Fig.\ 1a), in order to reduce the overall device resistance $R$. This shape is crucial because it has a small width $w$ in the central part of the channel -- ensuring small $A_{\rm active}$  and thus large $\Delta T$ -- while having a larger width towards the contacts -- minimizing the graphene sheet resistance. Furthermore, the large interface with the source and drain contacts minimizes the contribution of contact resistance to the overall resistance $R$. We assess the validity of the analytical model of Eqs. 1-2, by comparing the results with numerical simulations of the PTE photocurrent generated in different graphene geometries (see SI). We find agreement between the analytical and numerical results, showing the validity of our analytical approach. Importantly, the analytical model gives us insights into the physics that determines the detector response, thus allowing for optimization strategies. The advantage of the numerical simulations is that they are also valid for non-rectangular graphene shapes. Our detector design is the result of these analytical and numerical simulations.
\\

Based on the analytical model for the PTE response, we now examine our experimental results quantitatively. We have measured a photocurrent of $I_{\rm PTE}$ = 1.14 $\mu$A (Device A, 2.5 THz) for an incident power of $P_{\rm in}$ = 5 mW (focused to a spot size $A_{\rm focus}$, see Methods). Using Eq.\ \ref{PC} with $(S_1 - S_2)$ = 160 $\mu$V/K (estimated from Ref.\ \cite{dassarma09}) and $R$ = 2.3 k$\Omega$ (measured), we find an experimental temperature increase of $\sim$16 K (confirming the weak heating regime). Then using Eq.\ \ref{deltaT} with interfacial heat conductivity $\Gamma_{\rm cool}$ = 7$\cdot$10$^4$ W/m$^2$K (determined in Ref.\ \cite{Tielrooij2018}), cooling length $\ell_{\rm cool}$ = 510 nm (from the mobility and interfacial heat conductivity, see SI), and channel width $w$ = 2 $\mu$m (measured), we find the absorbed power (in the active area of the graphene channel) to be $P_{\rm abs}$ = 2.3 $\mu$W. We compare this value with the absorbed power we find from numerical simulations of the antenna-graphene structure, using the same irradiance as in the experiment (see Methods). These simulations give an absorbed power (in the entire graphene channel) of $P_{\rm abs, sim}$ = 7 $\mu$W (at 2.5 THz). This number is close to the number we obtained experimentally, adding credibility to our assignment of the PTE as the dominant photoresponse mechanism and to the validity of our analytical model. We ascribe the lower experimental value (by a factor $\sim$3) to non-optimal performance of the actual antenna in the photodetector device, most likely due to the presence of metallic regions around the antenna (see SI). Furthermore, the simulations consider the absorption in the entire graphene sheet, rather than only in the photoactive area of the graphene channel. Notably, we point out that without the antenna, the amount of incident THz light from a diffraction-limited spot that would be absorbed in the photoactive area of the graphene channel would be more than three orders of magnitude lower, highlighting the importance of the antenna-integration. We illustrate this in Fig.\ 3b, where we show the antenna-induced absorption enhancement, defined as $G = \frac{{P_{\rm abs, sim, w/\; antenna} }}{P_{{\rm abs, sim, w/o \; antenna} }}$. These simulations show that the antenna enhances the graphene absorption by more than three orders of magnitude.
\\

\newcolumntype{s}{>{\columncolor[HTML]{AAACED}} p{3cm}}
\setlength{\arrayrulewidth}{0.3mm}

\begin{savenotes}
\begin{table*}[ht]
\begin{threeparttable}
\centering 		
\begin{tabular}{|c|c|c|c|c|c|}			
\hline                       
\textbf{Reference} & \textbf{Mechanism} & \textbf{NEP} & \textbf{Normalization}  & \textbf{Speed} & \textbf{Freq. range} \\ [0.5ex]
& & \textbf{(pW/$\sqrt{\rm Hz}$)} & \textbf{area $A_{\rm norm}$}$^a$ & \textbf{(ns)} & \textbf{(THz)} \\
\hline			                  	
This work   & PTE & 80 & $\lambda^2/\pi $ & $<$30  & 1.8 -- 4.25  \\				

\hline
~\cite{fuhrer14}  & PTE & 1100 & $\sim\lambda^2$/3350 & 0.11 & 2.5 \\
\hline
~\cite{pte18}  & PTE & 350 & $\sim\lambda^2$/450 & 9000 & 0.08 -- 0.3 \\
\hline
~\cite{chalmers17}  & Rectification & 130 & none & - & 0.4 \\
\hline
~\cite{tredicucci14} & Plasma waves  & 2000 & $\lambda^2/4$ & - & 0.29 -- 0.38 \\ 
\hline
~\cite{bandurin18} & PTE/plasma waves  & 600 & none$^b$ & - & 0.13 -- 0.45 \\
\hline
~\cite{manchester17}   & Ballistic rectification & 34 & $\lambda^2/4\pi$ & - & 0.07-0.69 \\ [1ex]		    
\hline								
\end{tabular}
\caption{Comparison of graphene-based THz photodetectors.}				
\begin{tablenotes}
    \footnotesize
      \item $^a$ The normalization area $A_{\rm norm}$ refers to the area to which the incident power was normalized: $P_{\rm norm} = P_{\rm in}\cdot A_{\rm norm}/A_{\rm focus}$.
In our work, for example, we use the power in a diffraction-limited spot $P_{\rm diff} =  P_{\rm in}\cdot A_{\rm diff}/A_{\rm focus}$, $i.e.$\ we use $A_{\rm norm} = A_{\rm diff} = \lambda^2/\pi$. \\
$^b$ Whereas the incident power was not normalized to any area, a correction of the incident power was applied to account for losses occurring in the focusing system.  
   
    \end{tablenotes}
     \end{threeparttable}
\end{table*}
\end{savenotes}

Finally, we compare the performance of our photodetector with respect to other graphene-based THz photodetectors in the literature. We first compare with other detectors where the  photodetection mechanism was explicitly ascribed to the PTE effect (see Table 1). Since not every report used the same power normalization procedure for the responsivity and NEP, we mention explicitly the normalization procedure that was used. We note that our THz detector is 2--4 orders of magnitude more sensitive than any other THz PTE photodetector (if the same normalization procedure would be applied). We attribute this to our novel design with the antenna/gating structure, the optimal graphene channel geometry and the use of high-mobility hBN-encapsulated graphene. Furthermore, the sensitivity that we obtain is very similar to, or better than, the most sensitive graphene-based THz detectors reported in the literature~\cite{chalmers17, manchester17}. The operation of those detectors, however, has only been shown for frequencies below 1 THz and no response times have been measured. Additionally, it's important to point out that the Drude optical conductivity and therefore absorption in graphene is higher in the GHz range than in THz \cite{Mak2012c, Dawlaty2008b}, hence a direct comparison with detectors operating below 1 THz is not straightforward since we do not normalize the detector responsivity by the graphene absorption. 
\\

In conclusion, we have demonstrated a novel THz photodetector, which is dominated by the photo-thermoelectric effect. It operates at room temperature, is highly sensitive and very fast, has a wide dynamic range and operates over a broad range of THz frequencies. We have optimized the PTE THz detector by using a split-gate/antenna structure with narrow gap, which funnels the incident THz light exactly at the small photoactive area of the detector leading to strongly enhanced THz absorption in graphene. This structure simultaneously allows for tuning the detector to the optimal gating configuration, where a $pn$-junction is created in the graphene channel. Furthermore, we have used an ``H-shaped'', high-quality, hBN-encapsulated graphene channel with a narrow width, in order to have a small photo-active area, thus achieving a large THz-induced change in temperature, and a low overall device resistance. 
\\

Given the qualitative and quantitative understanding we have developed of the performance of our detector, we identify strategies for further improving its performance. Most importantly, by optimizing the antenna, a higher absorption and therefore lower NEP can be achieved. Additionally, by using a more broadband antenna, the detector will be sensitive for a larger range of THz frequencies. The sensitivity can be further enhanced by having a lower thermal conductivity $\Gamma_{\rm cool}$. This could be achieved by exploring alternative encapsulation materials, rather than hBN, $e.g.$\ a transition metal dichalcogenide (TMD) material, and by operating at a lower temperature. We estimate that it will be possible to reach an NEP in the low pW/$\sqrt{\rm Hz}$-regime. We expect that the unique combination of high sensitivity and fast operation means that these THz PTE detectors will play an important role in a large spectrum of applications. 
\\

\section{Supporting Information}
Raman, electrical characterization and optical images of the devices, fall time measurements of device A, power dependence and scanning photocurrent measurements of device B and thermoelectric simulations.

\section{Methods}

\subsubsection{Sample fabrication}
First we patterned the antenna/gate structure on transparent SiO$_2$ (Infrasil) substrate using electron beam lithography followed by evaporation of titanium (2 nm) / gold (30 nm).  The antenna gap was 200 nm (100 nm) for device A (B). We then released an hBN/graphene/hBN stack onto the antenna/gate structure. The stack elements (top and bottom hBN and graphene) were mechanically cleaved and exfoliated onto freshly cleaned Si/SiO$_2$ substrates. The full stack was prepared by the Van der Waals assembly technique \cite{Wang2013g,Pizzocchero2016} and released onto the antenna/gate structure. This was followed by patterning source and drain electrodes, using electron beam lithography with a PMMA 950 K resist film and exposing it to a plasma of CHF$_3$/O$_2$ gases for partially etching the stack. Consequently, we evaporate side contacts of chromium (5 nm) / gold (60 nm) and lift off in acetone as described in Ref.\ \cite{Wang2013g}{.} Finally, the encapsulated graphene was etched into the ``H'' shape using a plasma of CHF$_3$/O$_2$ gases. From gate-dependent measurements (varying $V_{\rm L}$ and $V_{\rm R}$ simultaneously) on Device A, we extract a mobility $>$20,000 cm$^2$/Vs and a contact resistance of 126 $\Omega$ (3.8 k$\Omega\cdot\mu$m) (see SI).

\subsubsection{Measurements}
In one setup (used for all Figs.\ except for Figs.\ 1d and 3c), we used a continuous wave THz beam from a gas laser (FIRL 100 from Edinburgh Instruments) providing a maximum output power in the range of a few tens of milliwatts, and frequencies of 1.83, 2.52, 3.11 and 4.25 THz. The device position was scanned using a motorized $xyz$-stage. The THz laser was modulated at 266 Hz using an optical chopper and the photocurrent was measured using a lock-in amplifier (Stanford). We verified that the output THz beam was strongly polarized (only 2\% of residual unpolarized light) and mounted the detector with the antenna axis parallel to the polarization. The THz light was focused using a gold parabolic mirror with focal distance 5 cm. The THz power was measured using a pyroelectric THz detector from Gentec-EO placed at the sample position. The second setup (used for Fig.\ 1d and 3c) contained a pulsed QCL at 3.4 THz, with an expected rise time $<$1 ns. 

\subsubsection{Responsivity calculation}
For the responsivity $\mathbb{R} = I_{\rm PTE}/P_{\rm diff}$, we extracted the PTE photocurrent $I_{\rm PTE}$ from the output signal of the lock-in amplifier $V_{\rm LIA}$ using $I_{\rm PTE} = \frac{2 \pi \sqrt{2}}{4G} V_{\rm LIA}$\cite{tredicucci12, tredicucci14}, where $G$ is the gain factor in V/A (given by the lock-in amplifier). The power in a diffraction-limited beam is given by $P_{\rm diff} = P_{\rm in}\cdot A_{\rm diff}/A_{\rm focus}$, where the ratio $A_{\rm diff}/A_{\rm focus} = \frac{w^2_{\rm 0,diff}}{w_{\rm 0,x} w_{\rm 0,y}}$. In order to obtain $w_{\rm 0,x}$ and $w_{\rm 0,y}$ we use our observation that the photoresponse is linear in laser power and measured the photocurrent while scanning the device in the $x-$ and $y-$direction. The photocurrent is then described by Gaussian distributions $\propto e^{-2x^2/w^2_{\rm 0,x}}$ and $\propto e^{-2y^2/w^2_{\rm 0,y}}$, where $w_{\rm 0,x}$ and $w_{\rm 0,y}$ are the respectively obtained spot sizes (related to the standard deviation via $\sigma = w_0/2$ and to the FWHM via FWHM = $\sqrt{2 \ln(2)}w_0$). For Device A, we obtained $w_{\rm 0,x}$ = 263.3 $\mu$m and $w_{\rm 0,y}$ = 331.2 $\mu$m. For the diffraction-limited spot, we took $w_{\rm 0,diff} = \frac{\lambda}{\pi}$, with $\lambda$ the THz laser wavelength. The diffraction-limited area is thus taken as $A_{\rm diff} = \pi w_{\rm 0, diff}^2 = \lambda^2/\pi$. 

\subsubsection{Antenna simulations}
The full wave simulations were performed in Comsol. The frequency dependent permittivity of hBN was taken from Ref. \cite{lundeberg17}{.} The optical conductivity of graphene was calculated employing the local random phase approximation at $T$ = 300 K with a scattering time of 100 fs. In the simulations, for simplicity, the Fermi energy of the graphene sheet was spatially constant. A plane wave source was used for illumination, where the incident power was normalized to give the same irradiance as the experiment. 

\subsubsection{Speed calculations}
The electrically-limited operation frequency of the detector is related to the RC-time constant $\tau = RC$, with $R$ total graphene resistance (including contact resistance), and $C$ total graphene capacitance. The operating speed is then given by the rate $f = (2\pi\tau)^{-1}$. The rise time $\tau_{\rm rise}$ is the measure of the photodetector response speed to a stepped light input signal. It is the time required for the photodetector to increase its output signal from 10\% to 90\% of the final steady-state output level. The rise time is calculated as $\tau_{\rm rise} = \tau\cdot \ln(9)$ = $(2\pi f)^{-1}\cdot \ln(9)$ =  0.35/$f$.

\section{Acknowledgments}
The authors thank Mark Lundeberg and Iacopo Torre for fruitful discussions. F.H.L.K. acknowledges financial support from the Spanish Ministry of Economy and Competitiveness, through the “Severo Ochoa” Programme for Centres of Excellence in R$\&$D (SEV-2015-0522), support by Fundacio Cellex Barcelona, Generalitat de Catalunya through the CERCA program,  and  the Agency for Management of University and Research Grants (AGAUR) 2017 SGR 1656.  Furthermore, the research leading to these results has received funding from the European Union Seventh Framework Programme under grant agreement no.785219 Graphene Flagship (Core2). ICN2 is supported by the Severo Ochoa program from Spanish MINECO (Grant No. SEV-2017-0706). K.-J.T. acknowledges support from a Mineco Young Investigator Grant (FIS2014-59639-JIN). S.C. acknowledges funding from the Barcelona Institute of Science and Technology (BIST), the Secretaria d’Universitats i Recerca del Departament d’Economia i Coneixement de la Generalitat de Catalunya and the European Social Fund – FEDER. 
M.S.V. acknowledges financial support from the ERC Project 681379 (SPRINT) and partial support from the second half of the Balzan Prize 2016 in applied photonics delivered to Federico Capasso. A.Y.N. acknowledges funding from the Spanish Ministry of Economy, Industry and Competitiveness, national project MAT2017-88358-C3-3-R.

\clearpage

\onecolumn
\newpage
 \section{Supporting Information}
 
\setcounter{figure}{0}

\makeatletter 
\renewcommand{\thefigure}{S\@arabic\c@figure}
\makeatother
\setcounter{table}{0}
\renewcommand{\thetable}{S\arabic{table}}
\makeatother

\section{1) Thermoelectric simulations}
To calculate the PTE response of our device we solved numerically the linearized thermoelectric equations~\cite{Landau1984}:
\begin{align}
\bm j_Q (\bm r) & = -\sigma(\bm r)\nabla V(\bm r) - \sigma (\bm r) S(\bm r)\nabla T(\bm r), \label{eq:current_charge}\\
\bm j_E(\bm r) & = -\Pi(\bm r)\sigma(\bm r)\nabla V(\bm r) - [\kappa(\bm r)+\Pi(\bm r)\sigma (\bm r) S(\bm r)]\nabla T(\bm r),
\label{eq:current_heat}
\end{align}
where $\bm j_Q (\bm r)$, $\bm j_E(\bm r)$ are the electric and energy current density respectively, $V(\bm r)$ is the voltage, $T(\bm r)$ is the local temperature, $\sigma(\bm r)$ is the electrical conductivity, $S(\bm r)$ is the Seebeck coefficient, $\Pi(\bm r)$ is the Peltier coefficient, and $\kappa(\bm r)$ is thermal conductivity. These two equations are coupled to the continuity equations for the charge and energy currents:
\begin{align}
\nabla \cdot \bm j_Q(\bm r) & = 0,\label{eq:continuity_charge}\\
\nabla \cdot \bm j_E(\bm r) & = - g(\bm r)[T(\bm r)-T_0]+J_E(\bm r). \label{eq:continuity_heat}
\end{align}
Here $g(\bm r)$ parametrizes the thermal conduction to the substrate~\cite{Tielrooij2018}, $T_0$ is the substrate temperature, and $J_E(\bm r)$ is a local heat source, that in our case is due to light absorption. We solved these equations inside a 2D simulation box $[0,L_{\rm tot}]\times[0,W_{\rm tot}]$ illustrated in Fig. S5 that includes the graphene sheet and the gold contacts.
We artificially included a small rectangular region of dimension $\delta \times W$ between the graphene and the gold contacts to account for finite contact resistance by choosing the value of its conductivity $\sigma=\delta/R_{\rm cont}$ with $R_{\rm cont}$ being the value of the gold-graphene contact resistance. We assumed that the material parameters $\sigma(\bm r)$, $S(\bm r)$, $\Pi(\bm r)$, $\kappa(\bm r)$, and $g(\bm r)$ are piecewise constant in the regions described in Fig. S5.
Since, because of Onsager relations, $\Pi(\bm r)=T_0 S(\bm r)$, and the thermal conductivity is related to the electrical conductivity by the Wiedemann-Franz law $\kappa(\bm r)=\mathcal{L}_0\sigma(\bm r)T_0$, with
$\mathcal{L}_0\equiv \pi^2 k_{\rm B}^2 /(3e^2)=2.44\cdot 10^{-8}{\rm W\Omega K^{-2}}$, we have only three independent parameters ($\sigma(\bm r), S(\bm r), g(\bm r)$) that are listed in table~\ref{mat_params}.

Eqs.~(\ref{eq:current_charge},\ref{eq:current_heat},\ref{eq:continuity_charge},\ref{eq:continuity_heat}) need to be supplemented with boundary conditions (BCs). 
We used Dirichlet BCs at $x=0,L_{\rm tot}$. These read:
\begin{align}\label{eq:dir_bc}
V(x=0,y) & =V_{\rm sd},\\
V(x=L_{\rm tot},y) & =0,\\
T(x=0,y) & =T(x=L_{\rm tot},y)=T_0.
\end{align}
We assumed instead homogeneous Neumann condition on the remaining part of the boundary
\begin{align}\label{eq:neu_bc}
\hat{\bm n}\cdot \bm j_Q (\bm r) & =0,\\
\hat{\bm n}\cdot \bm j_E (\bm r) & =0,  \label{eq:neu_bc_2}
\end{align}
$\hat{\bm n}$ being the outward normal unit vector. We solved numerically Eqs.~(\ref{eq:current_charge},\ref{eq:current_heat},\ref{eq:continuity_charge},\ref{eq:continuity_heat}), with BCs (\ref{eq:dir_bc},\ref{eq:neu_bc_2}) using the Finite Volume Method (FVM) on a regular square grid of 400x240 cells. Once the solution is found, the current flowing in the device can be calculated as

\begin{equation}
I\equiv \int_0^{W_{\rm tot}} J_x(0,y)dy=\int_0^{W_{\rm tot}} J_x(L_{\rm tot},y)dy=\frac{V_{\rm sd}}{R}+I_{\rm PTE},
\end{equation}

where the last equality holds because of linearity, $R$ is the resistance, and

\begin{equation}
I_{\rm PTE} = \int d{\bm r} \mathbb{R}_{\rm int}({\bm r}) J_E(\bm r).
\label{photocurrent_start}
\end{equation}
 
To find $R$ we simply calculate the current by setting $J_{\rm E}(\bm r)=0$. The responsivity $\mathbb{R}_{\rm int}({\bm r_0})$  at a given position $\bm r_0$ is instead calculated by setting $V_{\rm sd}=0$ and $J_{\rm E}=\delta(\bm r-\bm r_0)$, and calculating the corresponding current. This step is repeated for every $\bm r_0$ in the simulation grid to obtain the responsivity maps in Fig. S5.

\newcolumntype{s}{>{\columncolor[HTML]{AAACED}} p{3cm}}
\setlength{\arrayrulewidth}{0.3mm}

\begin{savenotes}
\begin{table*}[ht]
\begin{threeparttable}
\centering
\begin{tabular}{c|ccc}
Material & $\sigma [S]$ & $S [\mu V/K]$ & $\Gamma_{cool} [W/m^2K]$\\
\hline 
\hline
Graphene $n$ ($p$) doped & 1.3$\cdot$10$^{-3}$ $^a$ & 80 (-80)~\cite{dassarma09} & 7$\cdot10^4$ ~\cite{Tielrooij2018}\\
Gold & 4$^b$ & 0$^c$ & 300$\cdot10^4$~\cite{Cai2010, Koh2010a}\\
Interface contact & 2.5$\cdot$10$^{-6}$ $^d$ & 0$^c$ & 100$\cdot10^4$\\
\hline
\end{tabular}
\centering
\caption{\label{mat_params}{Material parameters}}				
\begin{tablenotes}
    \footnotesize
      \item $^a$ When considering E$_F$ = 50 (-50) meV, T = 300 K and 200 fs relaxation time for both doping type $n$ ($p$). $^b$ For 100 nm gold. $^c$ We neglect the contribution of the Seebeck coefficient of the metal. $^d$ For a $\delta$ = 25 nm.   
    \end{tablenotes}
     \end{threeparttable}
\end{table*}
\end{savenotes}
\clearpage
\newpage
\section{2) Supplementary figures}

\begin{figure} [H]
   \centering
    \onecolumn \includegraphics [scale=0.9]			
   {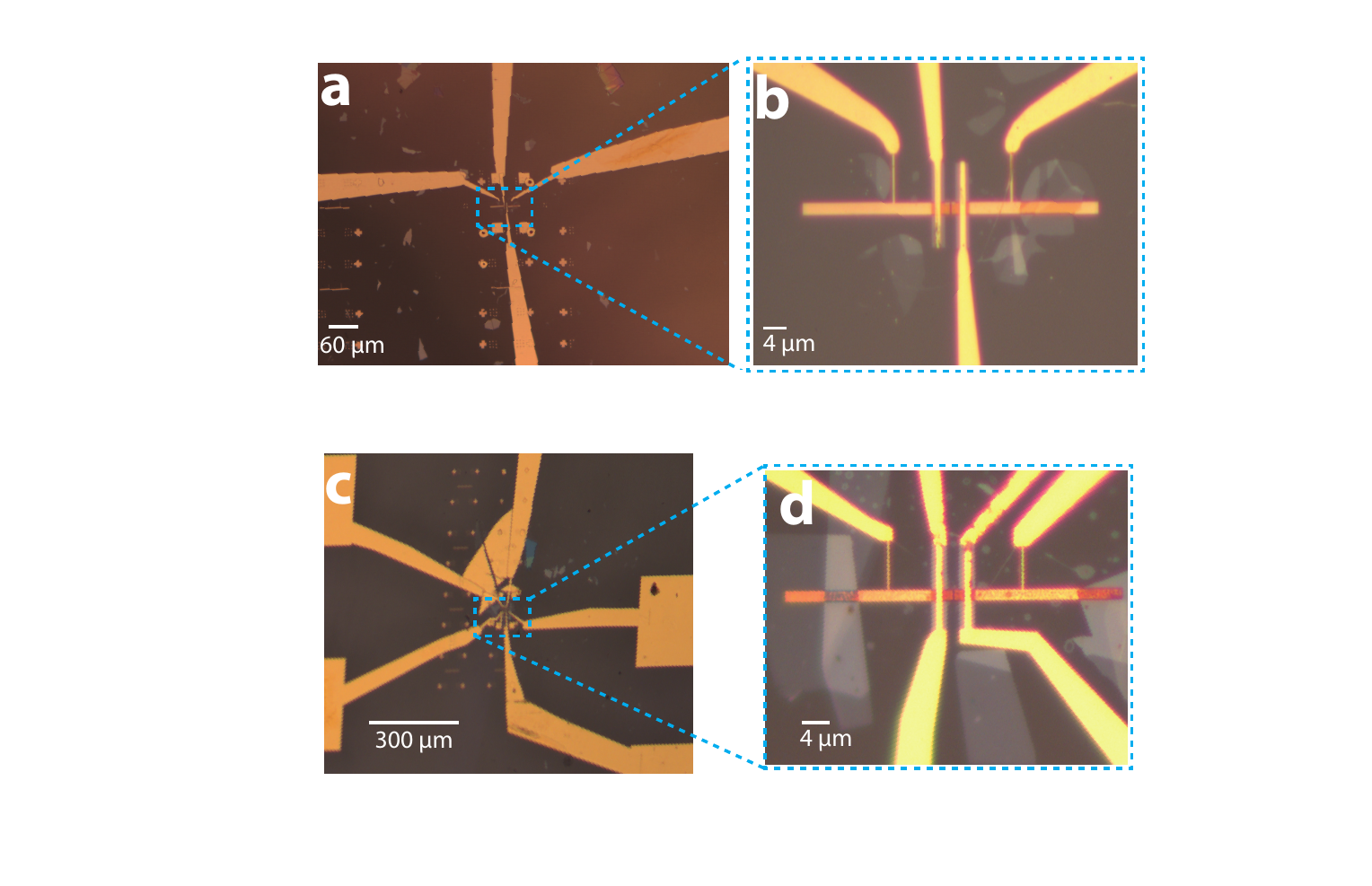}
   \caption{ 
Optical microscope images of the graphene-based THz photodetectors. \textbf{a)} 10x magnification image of Device A, showing the metallic rods contacting the antenna branches and source-drain electrodes. \textbf{b)} 100x magnification image of Device A, displaying the area of the photodetector containing the antenna/gate structure, source-drain electrodes and ``H-shaped'', h-BN encapsulated graphene. \textbf{c)} 5x magnification image of Device B and \textbf{d)} 100x magnification image of Device B.
}
\label{optical_pics}
   \end{figure}



\begin{figure*} [h!!!!!]
   \centering
   \includegraphics [scale=0.4]			
   {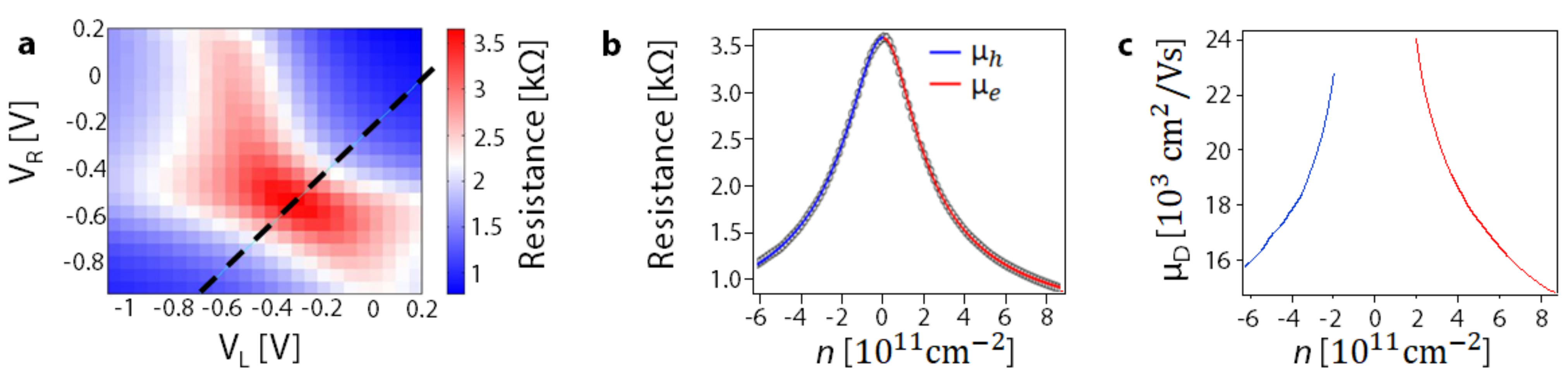}
   \caption{ 
\textbf{a)} Measured resistance map as a function of the right gate (left axis) and left gate (bottom axis) for Device A. \textbf{b)} Resistance as a function of graphene charge carrier density (linecut in dashed line shown in \textbf{a)}). We fit the resistance $R$ curve using the model from Ref.~\cite{kim07}{,} where $R = R_{\rm c} + (L/w)(\mu en)^{-1}$, $R_{\rm c}$ is contact resistance, $L$ is channel length, $w$ is channel width, $\mu$ is mobility, $e$ is elementary charge, and the carrier density is given by $n = \sqrt{(n^*)^2+(\beta(V_g - V_{\rm Dirac}))^2}$. Here $n^*$ is the residual doping concentration, $V_g$ is gate voltage, $V_{\rm Dirac}$ is the gate voltage that corresponds to the Dirac point, and $\beta = \epsilon_0\epsilon_{\rm hBN}/d_{\rm hBN}e$, where $\epsilon_0$ is vacuum permittivity, $\epsilon_{\rm hBN}$ the dielectric constant of hBN and $d_{\rm hBN}$ the thickness of the bottom hBN. We obtained 22,000 and 19,000 cm$^2$/Vs for electron (red) and hole (blue) mobility, respectively. We extracted a contact resistance of R$_c =$ 126 $\Omega$ (3.8 k$\Omega\cdot\mu$m), and a residual doping concentration of $n^* =$ 1.6$\cdot10^{11}$ cm$^{-2}$. \textbf{c)} Carrier mobility as a function of graphene charge carrier density extracted from the conductance measurement~\cite{Wang2013g}, where blue (red) corresponds to hole (electron) mobility.
}
\label{optical_pics}
   \end{figure*}



\begin{figure} [h!!!!!]
   \centering
   \includegraphics [scale=0.7]				
   {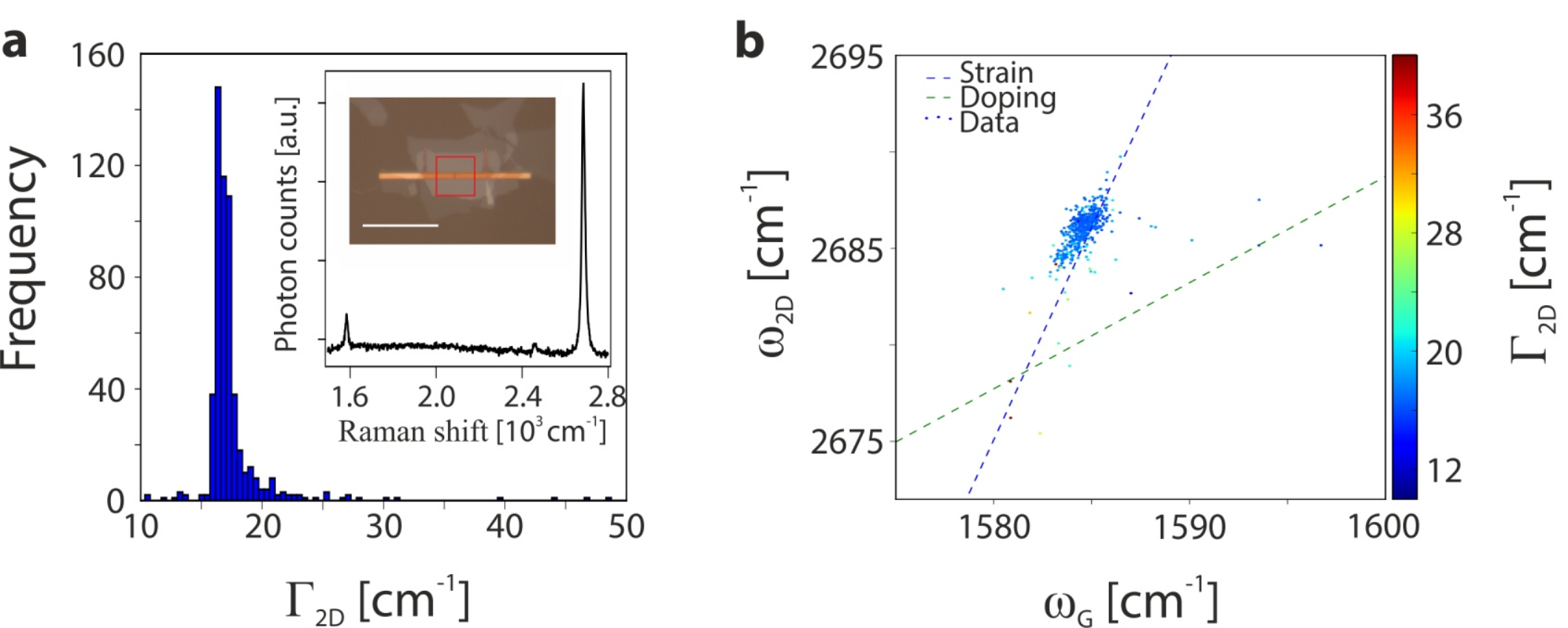}
   \caption{ 
Raman spectroscopy measurements of Device A. \textbf{a)} Histogram plot of the full-width-half-maximum of the graphene 2D peak ($\Gamma_{\rm 2D}$) across the $\sim$15x15 $\mu$m$^2$ region marked in red on the optical picture (see inset). The white scale bar corresponds to 30 $\mu$m. The inset shows a typical Raman spectrum of the monolayer graphene. The mean $\Gamma_{\rm 2D}$ is $\sim$17.5 cm$^{-1}$, characteristic of high quality single layer graphene~\cite{Lee2012d}. \textbf{b)} 2D-peak versus G-peak frequency ($\omega_{\rm 2D}$ and $\omega_{\rm G}$, respectively) extracted from the same Raman map as in panel \textbf{a}. The colorbar represents the $\Gamma_{\rm 2D}$ of the recorded spectrum. The data shows low doping, as confirmed in transport measurements (see Fig. S2), and moderate levels of strain~\cite{Lee2012d}.
}
\label{raman}
   \end{figure}



\begin{figure} [h!!!!!]
   \centering
   \includegraphics [scale=0.45]		
   {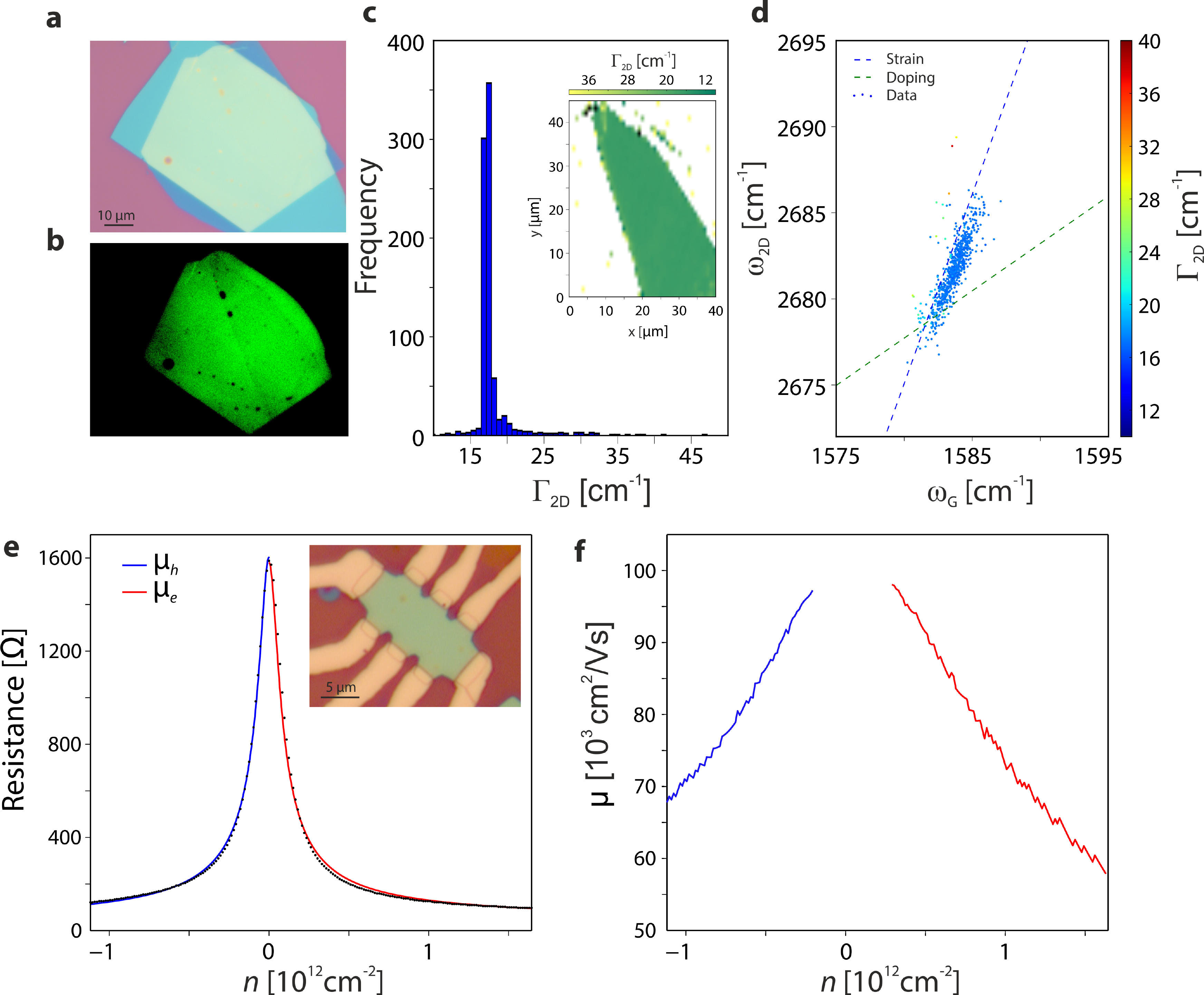}
   \caption{ 
Characterization of the Hall bar reference device fabricated following the same procedure as the photodetector (see Methods). \textbf{a)} Optical image of the initial hBN/graphene/hBN stack. \textbf{b)} Optical image shown in panel \textbf{a} with enhanced contrast. The graphene flake is optically visible (compare the graphene shape with the Raman map shown in the inset of panel \textbf{c}). \textbf{c)} Histogram plot of the Raman scan (see inset) showing the full-width-half-maximum of the graphene 2D Raman peak ($\Gamma_{\rm 2D}$). The mean $\Gamma_{\rm 2D}$ is $\sim$ 18.1 cm$^{-1}$, characteristic of high quality single layer graphene. \textbf{d)} Scattering plot of the 2D-peak versus the G-peak frequency ($\omega_{\rm 2D}$ and $\omega_{\rm G}$, respectively), where the colorbar represents the $\Gamma_{\rm 2D}$ of the recorded spectrum (inset of panel \textbf{c}). The Raman data shows comparable behavior as the photodetector (see Fig.\ S3b) with low $\Gamma_{\rm 2D}$, low doping and moderate levels of strain~\cite{Lee2012d}.
\textbf{e)} Resistance as a function of the charge carrier density of the measured Hall bar device (see inset). We fit the resistance $R$ curve using the model from Ref.~\cite{kim07}{,} and obtained 103,000 and 96,000 cm$^2$/Vs for the electron (red) and hole (blue) mobilities, respectively. \textbf{f)} Carrier mobility as a function of the charge carrier density extracted from the conductance measurement~\cite{Wang2013g}{,} where blue (red) corresponds to hole (electron) mobility. These results show that a mobility of 100,000 cm$^2$/Vs for these kind of photodetector devices is realistic, and therefore a detection speed in the 10 ps-range can be achieved. 
}
\label{optical_pics}
   \end{figure}



\begin{figure} [h!!!!!]
   \centering
   \includegraphics [scale=0.48]			
   {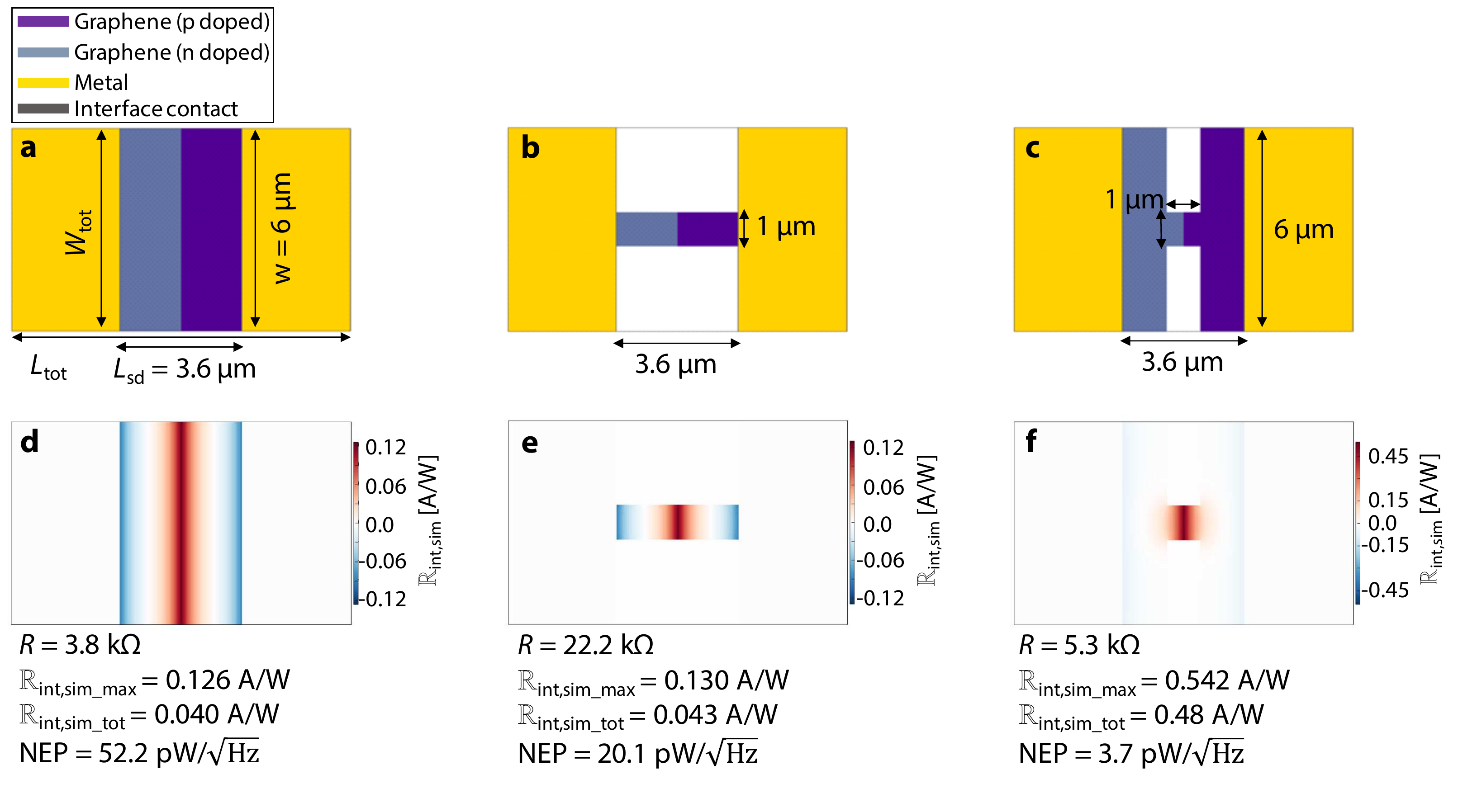}
   \caption{ 
Thermoelectric simulations results~\cite{lundeberg18}, as in Fig.\ 1c, which represent the resulting photoresponse after local photoexcitation, expressed in photocurrent normalized by absorbed power, $i.e.$\ internal responsivity ($\mathbb{R_{\rm int,sim}}$ as defined in Eq. S11, where $J_E(\bm r)$ represents the locally absorbed power). These simulations were based on similar characteristics of the measured samples, with a graphene scattering time of 200 fs (mobility 20,000 cm$^2$/Vs), contact resistance of 10 k$\Omega\cdot\mu$m, 3.6 $\mu$m of source-drain distance ($L_{sd}$), $\Gamma_{\rm cool}$ = 7$\cdot$10$^4$ W/m$^2$K, E$_F$ = 50 meV and -50 meV for $n$- and $p$-doped graphene regions respectively. Panels \textbf{a-c)} show the simulated device geometries and panels  \textbf{d-f)} the internal responsivity ($\mathbb{R_{\rm int,sim}}$) calculated at each position across the device. The $R$, $\mathbb{R_{\rm int,sim\_max}}$, $\mathbb{R_{\rm int,sim\_tot}}$ and NEP below panels \textbf{d-f)} indicate total device resistance (including graphene sheet and contact resistance), the maximum internal responsivity generated at the $pn$ junction, total internal responsivity (which takes into account also the opposite sign responsivity generated at the metal-graphene interface, i.e. $\mathbb{R_{\rm int,sim\_tot}} = \mathbb{R_{\rm int,sim\_max}}-\mid\mathbb{R_{\rm int,sim\_min}}\mid$) and noise-equivalent power, respectively. The simulations show that decreasing the width of the graphene channel (going from the design of panel \textbf{a} to the design of panel \textbf{b}) does not lead to any change in responsivity, but a decrease in noise-equivalent power. The responsivity is constant because of the trade-off between the increased $\Delta T$ due to the smaller active area and the increased device resistance due to the smaller width. The NEP is reduced, because the higher device resistance leads to reduced thermal noise. Then, by using the design with an ``H-shaped'' graphene channel in panel \textbf{c}, we both increase responsivity and decrease NEP. The main reason is that the active area is significantly reduced, without increasing too much the device resistance. Furthermore, the maximum responsivity and total internal responsivity are now very similar, because the PTE response at the $pn$-junction dominates. Therefore, this is the preferred design. We compare the simulation results for the internal responsivity $\mathbb{R_{\rm int,sim}}$ with the analytically obtained values ($\mathbb{R_{\rm int,analyt}}$) using the equations in the main text and the device resistances obtained from these simulations, and obtain $\mathbb{R_{\rm int,analyt}} = $ 0.12, 0.13 and 0.55 A/W for the designs from panels \textbf{a)}, \textbf{b)} and \textbf{c)} respectively. These analytical values are in excellent agreement with the $\mathbb{R_{\rm int,sim\_max}}$ values obtained from simulations. 
}
\label{optical_pics}
   \end{figure}



\begin{figure} [h!!!!!]
   \centering
   \includegraphics [scale=0.8]
   {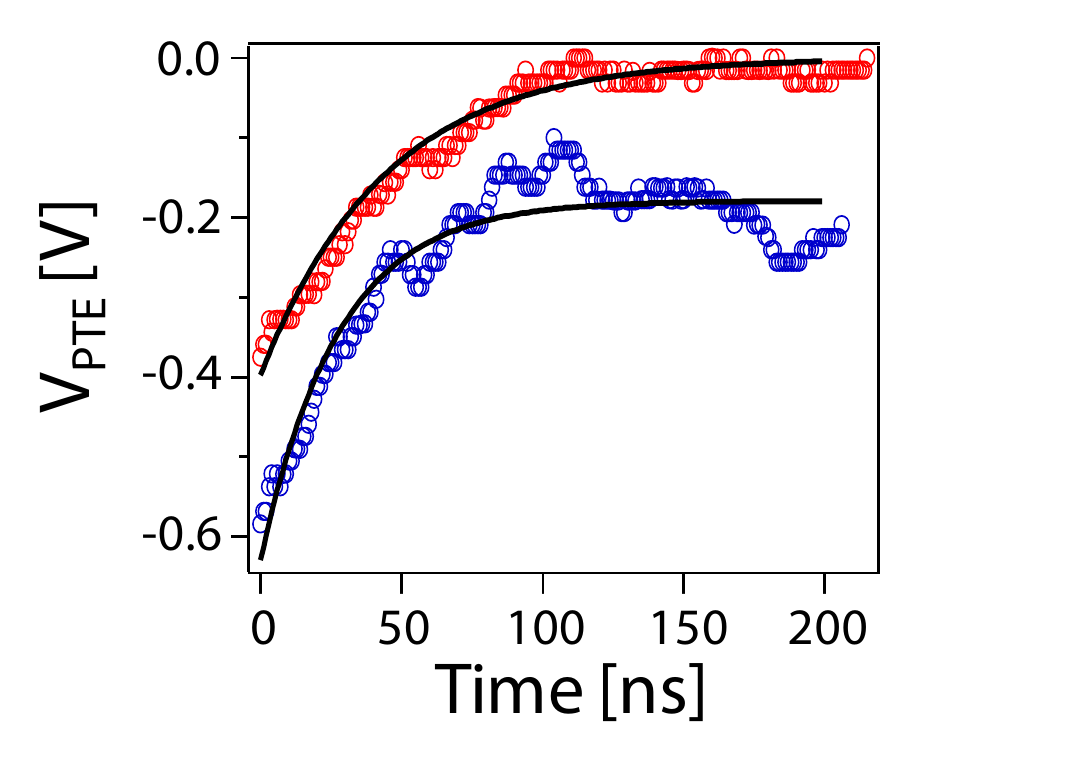}			
   \caption{ 
Results of the pulsed laser experiment, showing the fall time corresponding to the two regions (red and blue, the latter one plotted with an offset) of the inset of Fig.\ 3c of the main text. The open dots show the acquired photovoltage and the black lines represent the fit results with the exponential fits. We obtain an exponential fall time of 47 (28) ns for the red (blue) curve. As in the case of the rise time determination in the main text, this timescale is limited by the current amplifier with a bandwidth of 3.5 MHz.
}
\label{speed}
   \end{figure}


\begin{figure} [h!!!!!]
   \centering
   \includegraphics [scale=0.8]			
   {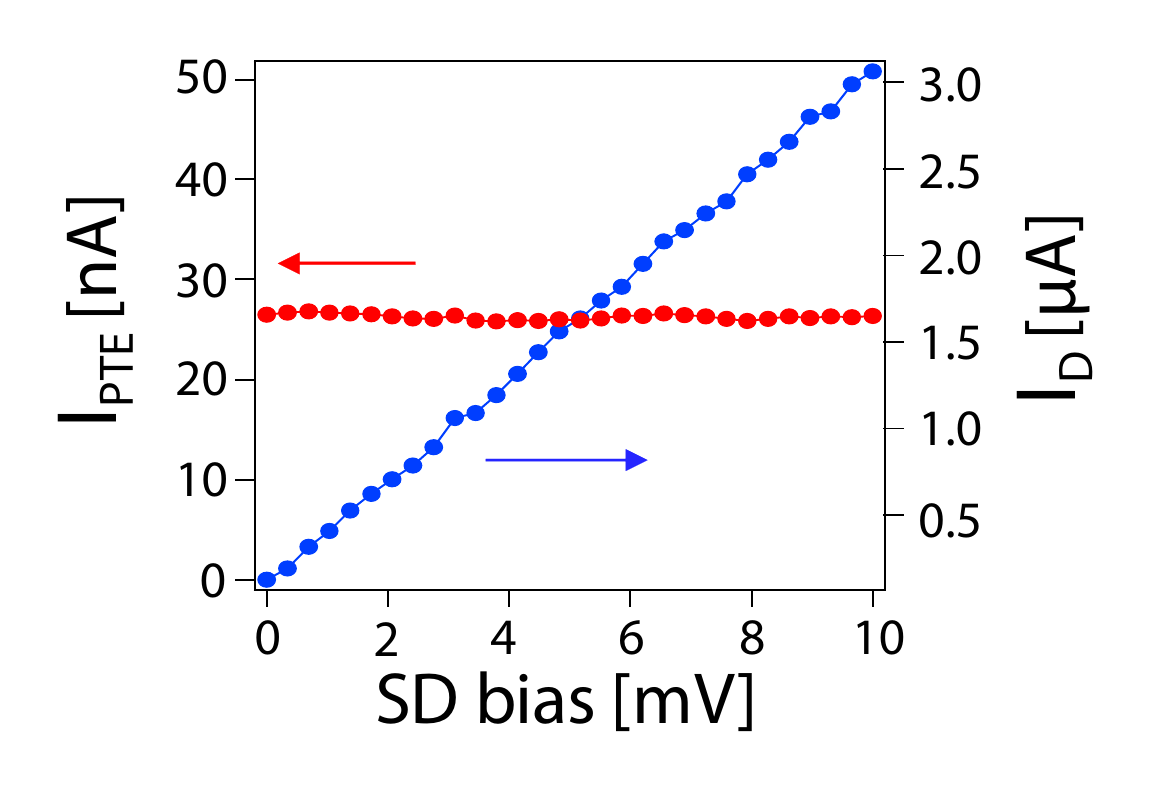}
   \caption{ 
Photocurrent and drain current measured simultaneously as a function of applied bias voltage between the source and drain contacts. We notice a linear increase of the drain current while increasing the bias, whereas the photocurrent remains constant.
}
\label{speed}
   \end{figure}



\begin{figure} [h!!!!!]
   \centering
   \includegraphics [scale=1.2]
   {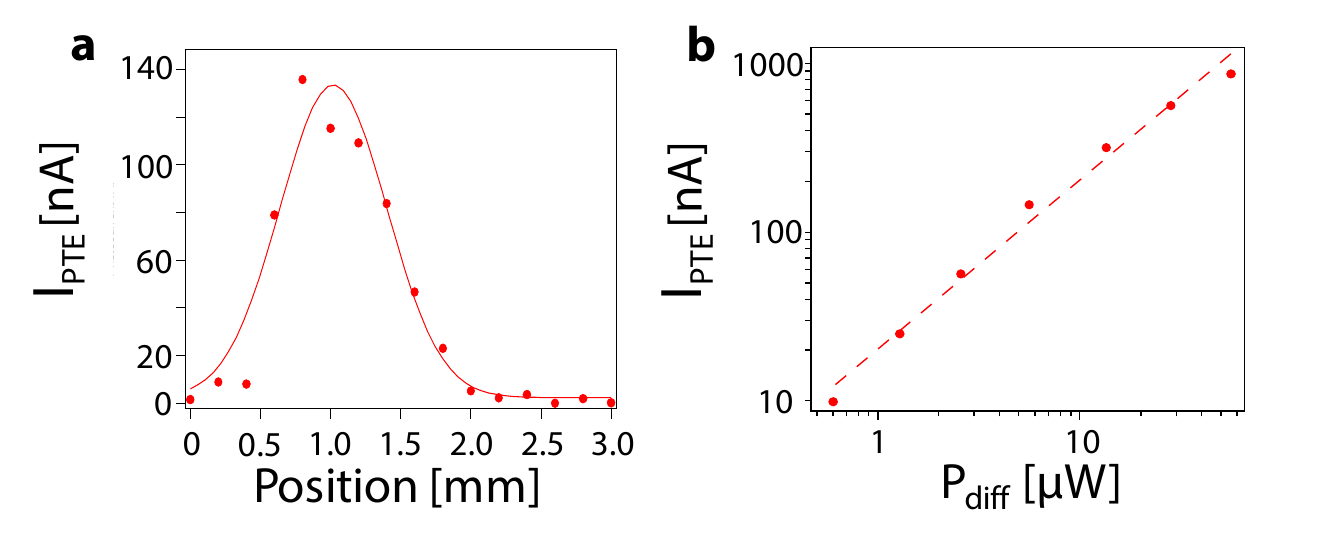}
   \caption{ 
Characterization of Device B. \textbf{a)} Photocurrent generated as a function of the scan position of the device. The red circles correspond to the experimental points and the red line to the Gaussian distribution fit. We obtain a beam focus FWHM of 897 $\mu$m at 2.52 THz. We observe a maximum $I_{\rm PTE}$ of 136 nA for an incident power ($P_{\rm in}$) of 2.88 mW ($P_{\rm diff} = P_{\rm in}\cdot A_{\rm diff}/A_{\rm focus}$ = 7.1 $\mu$W). Thus, the responsivity is $\mathbb{R} = I_{\rm PTE}/P_{\rm diff}$ = 19 mA/W (80.4 V/W). The sample was electrostatically doped with the left ($n$-doped, 60 meV) and right gate ($p$-doped, 50 meV) forming a $pn$-junction. \textbf{b)} Photocurrent as a function of incident power. It shows a linear trend over three orders of magnitude according to the fit displayed in dashed line. The maximum responsivity was $\mathbb{R}$ = 25 mA/W (105 V/W), from a photocurrent of  $I_{\rm PTE}$ = 145 nA, for an incident power of $P_{\rm in}$ = 2.28 mW ($P_{\rm diff}$ = 5.6 $\mu$W, for the same beam focus at 2.5 THz and $pn$-junction configuration as in panel \textbf{a)}. Using this responsivity and the Johnson noise that corresponds to the 4.2 k$\Omega$ measured device resistance, we find an NEP of 80 pW/$\sqrt{\rm Hz}$.
}
\label{speed}
\end{figure}



\begin{figure} [h!!!!!]
   \centering
   \includegraphics [scale=0.8]				
   {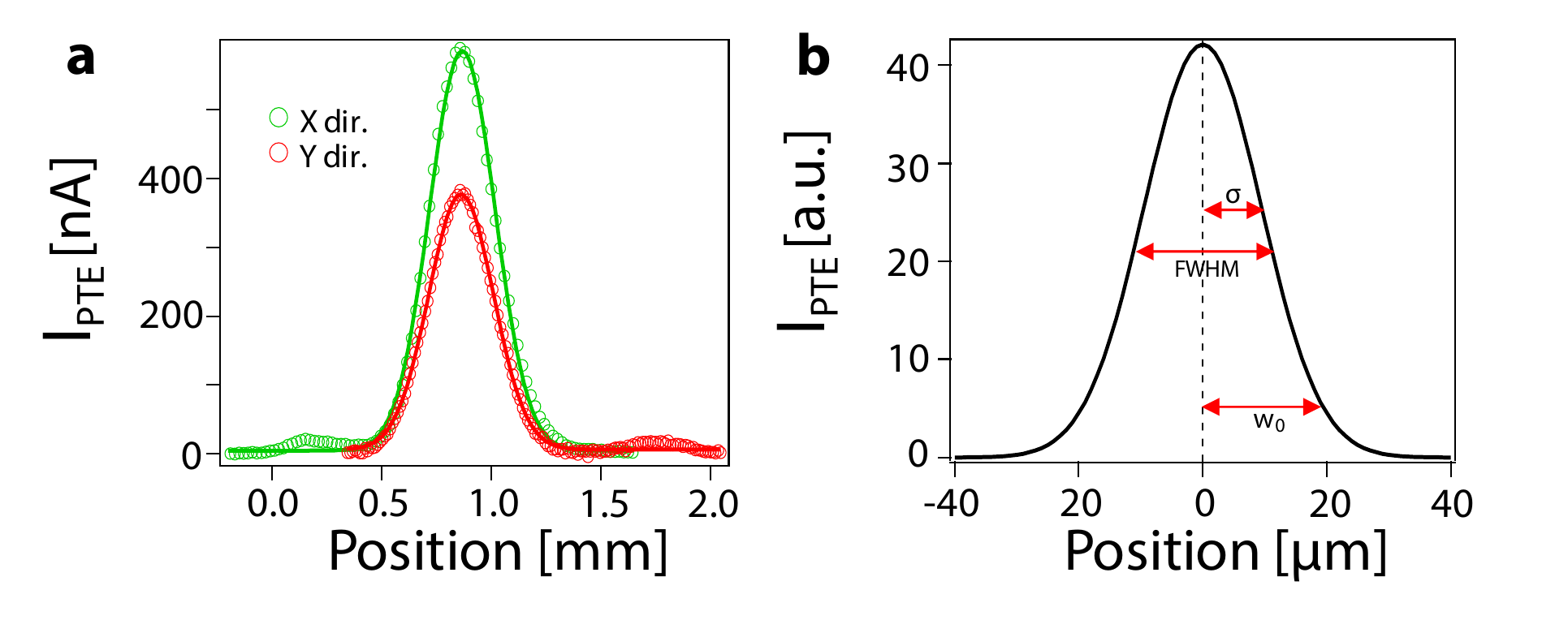}
   \caption{ 
Characterization of spot size for measurements on Device A. \textbf{a)} Photocurrent as a function of the scan position in $x$-direction (green) and $y$-direction (red) of the motorized stage onto which the device is mounted. The laser was tuned to 2.52 THz. The open dots represent the experimental data and the lines the fits according to Gaussian distributions. We obtain a FWHM of 351 (348) $\mu$m for $x$- ($y-$) scan direction. \textbf{b)} Calculated Gaussian distribution representing a diffraction-limit focus at 2.52 THz. The arrows indicate the standard deviation $\sigma$, FWHM and spot size $w_0$, corresponding to 60, 50 and 13.5$\%$ of the maximum, respectively. 
}
\label{speed}
\end{figure}



\begin{figure} [h!!!!!]
   \centering
   \includegraphics [scale=0.5]			
   {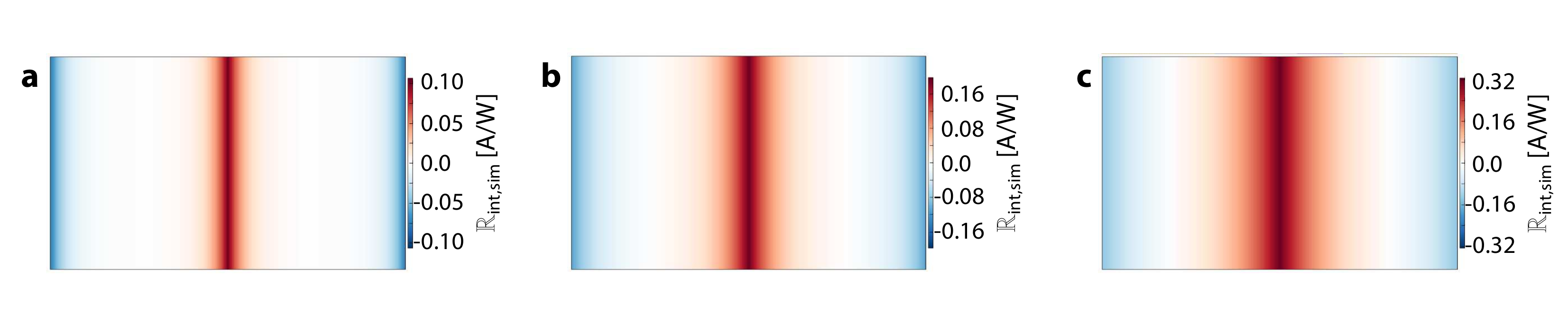}
   \caption{ 
Thermoelectric simulations~\cite{lundeberg18}{,} showing the internal responsivity ($\mathbb{R_{\rm int,sim}}$ as defined in Eq. S11, where $J_E(\bm r)$ represents the locally absorbed power) for a fixed device geometry and three different values of the interfacial heat conductivity $\Gamma_{\rm cool}$. For all these three cases the device length and width dimensions are 10 and 6 $\mu$m respectively, E$_F$ = 50 meV (-50 meV) for $n$- ($p$-) doped region and with a contact resistance of 10 k$\Omega\cdot\mu$m. The $\Gamma_{\rm cool}$ is 7$\cdot$10$^4$, 2$\cdot$10$^4$ and 0.7$\cdot$10$^4$ W/m$^2$K for \textbf{a)}, \textbf{b)} and \textbf{c)} respectively. As predicted from Eq.\ 2 in the main text, when decreasing $\Gamma_{\rm cool}$, $\Delta T$ increases and therefore $\mathbb{R_{\rm int,analyt}}$. In fact, by calculating the $\mathbb{R_{\rm int,analyt}} = I_{\rm PTE}/P_{\rm abs} = \frac{\Delta S \Delta T}{P_{\rm abs}R} = \frac{\Delta S}{A_{\rm active}\Gamma_{\rm cool}R}$ as shown in the main text, we obtain excellent agreement between these calculations and the thermoelectric simulations values of $\mathbb{R_{\rm int,sim}}$ for all three cases, namely $\mathbb{R_{\rm int,analyt}}$ = 0.10, 0.19 and 0.33 A/W for \textbf{a)}, \textbf{b)} and \textbf{c)} respectively. These results show that the responsivity increases (and the NEP decreases) with the square-root of the decrease in interfacial heat conductivity. 
}
\label{speed}
\end{figure}

\clearpage

\newpage


\bibliography{ThzPaper}

\clearpage

\end{document}